\documentclass{elsarticle}

\usepackage{amssymb}
\usepackage{amsthm}
\usepackage{amsfonts}
\usepackage{amsmath}
\newcommand*\rfrac[2]{{}^{#1}\!/_{#2}}

\usepackage{url}
\usepackage{tikz}
\usepackage{pgfplots}
\pgfplotsset{compat=1.15}
\usepackage{caption}

\usepackage{subcaption}
\newcommand\myrowlabel[1]{
  \rotatebox[origin=c]{90}{#1}
}
\usepackage{lipsum}
\usepackage{enumitem}
\usepackage[colorlinks]{hyperref}

\usetikzlibrary{shapes.geometric, arrows}
\tikzstyle{startstop} = [rectangle, rounded corners, minimum width=3cm, minimum height=1cm,text centered, draw=black, fill=red!30]
\tikzstyle{io} = [trapezium, trapezium left angle=70, trapezium right angle=110, minimum width=1.5cm, minimum height=0.5cm, text centered, draw=black, fill=blue!30]
\tikzstyle{process} = [rectangle, minimum width=2.8cm, minimum height=1cm, text centered, text width=2cm, draw=black, fill=orange!30]
\tikzstyle{process2} = [rectangle, minimum width=3cm, minimum height=1cm, text centered, text width=2cm, draw=black, fill=gray!30]
\tikzstyle{decision} = [diamond, minimum width=1.5cm, minimum height=0.5cm, text centered, text width=2cm, draw=black, fill=green!30]
\tikzstyle{decision0} = [diamond, minimum width=1.5cm, minimum height=0.5cm, text centered, text width=2cm, draw=black, fill=red!30]
\tikzstyle{arrow} = [thick,->,>=stealth]

\usepackage{graphicx}
\graphicspath{{figures/}}
\DeclareGraphicsExtensions{.pdf,.jpeg,.png,.jpg,.svg}

\biboptions{comma,round}

\begin{document}

\begin{frontmatter}




\title{Sub-band Coding of Hexagonal Images}

\author[label1]{Md Mamunur Rashid}
\author[label2]{Usman R. Alim}
\address[label1]{Seequent, Calgary AB, Canada \\ mamun.rashid@seequent.com}
\address[label2]{Department of Computer Science, University of Calgary, Calgary AB, Canada\\
ualim@ucalgary.ca}



\begin{abstract}
According to the circle-packing theorem, the packing efficiency of a hexagonal lattice is higher than an equivalent square tessellation. Consequently, in several contexts, hexagonally sampled images compared to their Cartesian counterparts are better at preserving information content. In this paper, novel mapping techniques alongside the wavelet compression scheme are presented for hexagonal images. Specifically, we introduce two tree-based coding schemes, referred to as SBHex (spirally-mapped branch-coding for hexagonal images) and BBHex (breadth-first block-coding for hexagonal images). Both of these coding schemes respect the geometry of the hexagonal lattice and yield better compression results. Our empirical results show that the proposed algorithms for hexagonal images produce better reconstruction quality at low bits per pixel representations compared to the tree-based coding counterparts for the Cartesian grid.

\end{abstract}


\end{frontmatter}

\section{Introduction}
\label{intro}
An image is a matrix of pixels arranged in a Cartesian grid. However, we are fundamentally interested in processing hexagonally sampled images that have been discretized on an appropriate hexagonal grid. The hexagonal lattice has several advantages over the Cartesian grid such as efficient sampling density, consistent connectivity, greater angular resolution and reduced aliasing. 
Despite these advantages, the hexagonal lattice is rarely used in practice as the data pipelines for hexagonal images lack appropriate tools for hexagonal image processing. Image acquisition devices are usually designed to capture images on a Cartesian grid. As a consequence, subsequent steps that deal with compression, transmission and display of these images are also designed to work with Cartesian sampled images. 

One of the key ingredients in hexagonal image processing is a suitable compression scheme. This paper focuses on hexagonal image compression using tree-based sub-band coding techniques. While similar techniques for Cartesian sampled images have been known for decades and are widespread, the equivalent coding techniques for hexagonal images have received very little attention. In particular, in this paper, we propose novel traversal scan orders and mapping techniques for spatially oriented trees for tree-based sub-band coding of hexagonal images. The results are compared both quantitatively and qualitatively with an equivalent tree-based algorithm for Cartesian images namely Embedded Zerotree Wavelet (EZW)~\cite{Shapiro.1993}.
Our main motivation is to demonstrate the advantages of hexagonal geometry in the context of sub-band image coding. We investigate the space saving and subject fidelity of images when the geometry of the hexagonal grid is respected during encoding. After implementing and exploring several mapping schemes and scan orders in different scenarios, we find that if we use a compatible parent-to-child relationship and block arrangement aided by a hexagonal scan order which follows the orientation of the hexagonal grid structure, more compression is possible compared to the traditional parent-to-child relationships for Cartesian grids associated with the raster scan or Morton scan~\cite{MOR66}. 

Our starting point is a wavelet representation of a hexagonally sampled image. Although there are a few choices of hexagonal wavelets available (see e.g.~\cite{condat07,fujinoki09}), we rely on the second order wavelet proposed by Cohen and Schlenker~\cite{Cohen1993} as it is dyadic, biorthogonal and compactly supported. Our goal in this work is to investigate a suitable encoding scheme that can be used in conjunction with this wavelet. Given the hexagonal wavelet coefficients, we propose two tree-based coding schemes, referred to as SBHex (Spirally-mapped Branch-coding for Hexagonal images) and BBHex (Breadth-first Block-coding for Hexagonal images). Our results show that our proposed schemes for hexagonal images produce better reconstruction quality at lower bits-per-pixel (BPP) values compared to the tree-based coding counterpart for the Cartesian grid that is based on a comparable Cartesian wavelet. Our algorithms (BBHex and SBHex) generally produce compressed files that are $2$ to $2.5$ times smaller than their Cartesian counterpart on datasets of generic images across all quality levels when BPP $< 4$. 

Our hexagonal coders are inspired by techniques found in tree-based encoding techniques for Cartesian images. In particular, we adapt EZW for use with hexagonal wavelet coefficients in a manner that respects the isotropic geometry of the hexagonal lattice. Our specific modifications are as follows. 
\begin{enumerate} [label=\roman*., itemsep=0pt, topsep=0pt] 
\item Incorporation of the discrete hexagonal wavelet transform (DHWT) aided by an index mapping technique.
\item Spiral branch-mapping of the \emph{hexagonal} wavelet tree which groups similar spatially located coefficients at different levels of detail into corresponding regions according to directionality. 
\item Hexagonal traversal order which scans the spiral wavelet tree according to the hexagonal geometry of the lattice. 



\item Helical parent-to-children relationships expressly for the spiral wavelet tree.

\item Breadth-first-traversed block-based coding of sub-trees using the proposed parent-to-children relationship which encodes the tree remarkably faster.
\end{enumerate}
\vspace{0.5em}

There are several steps involved in a production image compression pipeline such as JPEG2000~\cite{rabbani2002} and adapting all of these steps for a hexagonal grid is beyond the scope of this work. While they are both important considerations, we do not investigate the effect of using other types of hexagonal wavelets or alternate ways of entropy coding (we using Huffmann coding in this work). Our hope is to establish a baseline by studying in detail the sub-band coding scheme used to code the hexagonal wavelet coefficients. We also compare our results with JPEG2000 to assess the overall performance compared to the state of the art. In future work, we plan to improve upon this baseline by investigating other components of the compression pipeline. 

The remainder of the paper is organized as follows. In Section~\ref{sec:related}, influential research work related to image compression on both Cartesian and hexagonal grids is discussed. Additional relevant background pertaining to hexagonal sampling and wavelet decomposition is provided in Section~\ref{sec:background}. Different parts of our proposed sub-band coding schemes are described in Sections~\ref{sec:SBHex} (SBHex) and~\ref{sec:bbhex} (BBHex). Section~\ref{sec:result_discussions} presents and analyzes our experimental results. Finally, we conclude the paper in Section~\ref{sec:conclusion} summarizing the work and discussing some of the future directions. 

\section{Related work}
\label{sec:related}
\subsection{Square Images}
There is a wide and diverse array of compression algorithms for square pixelated tessellations. 
During 1985--2000, a number of transform-based image compression techniques emerged.
These were inspired by the discrete cosine transform (DCT). The ultimate result of these collaborative efforts was the standard ITU-T Rec. T-81— ISO/IEC 10918-1 developed by the Joint Photographic Experts Group (JPEG)~\cite{WallaceG.K1992TJsp}. The standard is considered the first ubiquitous coding standard for grayscale and color images~\cite{pennebaker_mitchell_2004}. It divides a source image into $8\times8$ blocks which are transformed with DCT independently. In lossless mode, encoding is executed by a variant of run-length symbols and Huffman coding.
One of the key constraints of the JPEG standard is the implementation of distinct modes for lossless and lossy compression which are independent of each other. Moreover, it also has separate progressive and hierarchical modes. JPEG2000~\cite{rabbani2002} addresses this constraint and also provides many other features such as region-of-interest coding and multiple levels of resolution.



During the development of JPEG, sub-band (or wavelet) coding systems also received considerable attention~\cite{WoodsJ1986Scoi}.
These systems are generally classified into two categories: block-based and tree-based. Block-based systems encode the sub-bands individually and independently. The bit rate for each sub-band is assigned based on its variance. 
The block coding is independently performed on image tiles (non-overlapping blocks) within individual sub-bands. The code blocks are encoded starting from the most significant bit (MSB) plane to bit planes of lower significance. Encoding of each bit plane consists of three passes: significance propagation, magnitude refinement, and cleanup. Tree-based systems encode spatial oriented trees (SOTs), or zero-trees. A zero-tree consists of all the nodes inside the SOT which are marked as insignificant. Number of coordinate locations in SOTs varies based on the root node and the parent-to-child relationship. The embedded zerotree wavelet (EZW) algorithm~\cite{Shapiro.1993} and the set partitioning in hierarchical trees (SPIHT) algorithm~\cite{pearlman_said_2011} are two well-known tree-based coders which provide the scaffolding for our work.

The current trend in image compression research is to use approaches based on machine learning (ML). 
Contemporary ML techniques for image compression employ a different variant of neural networks for image compression. The evolution and historical development of the neural network-based compression methodologies are mainly comprised of the multilayer perceptron~\cite{NampholA1996Icwa}, random neural networks~\cite{GelenbeE1994Rnla, HaiF2001Vcww}, convolutional neural networks~\cite{waveone2017, CavigelliLukas2017CAdc_cas_cnn, MentzerFabian2018CPMf}, recurrent neural networks~\cite{TodericiGeorge2015VRIC, TodericiGeorge2016FRIC, MinnenDavid2018Saic}, and finally generative adversarial networks~\cite{NIPS2014_5423, AgustssonEirikur2018GANf}. At a high level, these approaches attempt to train a model to minimize a loss function which attempts to match the training data as closely as possible. Despite showing some promising results, ML-approaches are constrained by high computational resources, excessive consumption of memory, power, and dependency on large training data sets~\cite{MaSiwei2019IaVC}.

\subsection{Hexagonal Images}
In comparison to Cartesian images, hexagonal image compression has received surprisingly limited attention.
Wang~\textit{et al.}~\cite{HuaqingWang2006ANAf, WangHuaqing2008FICo} explored the possibilities of fractal image compression on hexagonal images. The underlying pipeline consists of the transformation of images on a rectangular grid to a virtual spiral grid and vice-versa. At the heart of the process, straight-forward fractal coding is executed on the spiral planning of the image. Using a full search scheme, the encoding of hexagonal-based compression is computationally intensive. Ong and Fan~\cite{Ghim2007} assayed to reduce the time complexity by parallelizing the non-overlapped routines. However, to counter a large computational load, high computing resources are required to speed up the reconstruction.
Mang~\textit{et al.}~\cite{SilinMang2010CICB} and Jeevan~\cite{jeevan2012} proposed DCT-based compression approaches for hexagonal images. Both of the works initiate with hexagonal resampling from the rectangular grid succeeded by a hexagonal discrete cosine transform. Mang \textit{et al.}~\cite{SilinMang2010CICB} performed the compression doing vector quantization and associated entropy coding (not described in the paper). In Jeevan's work~\cite{jeevan2012}, compression happens based on the alternate pixel suppressal method where alternate rows and columns are discarded at each pass. Truhans~\cite{Aleksejs:2012} proposed DWT-based lossless compression based on a hexagonal to orthogonal grid conversion known as H2O~\cite{CondatL2007HRHG}. The chosen multiresolution filter bank is based on the work of Cohen and Schlenker~\cite{Cohen1993}. From the paper, it is not apparent how the entropy coding was formulated. Jeevan and Krishnakumar~\cite{JeevanK.M2014Coir} also worked on DWT-based compression where sampling quality was improved using a Gabor filter~\cite{jeevan2012}. The portion of DWT is not clearly described in their work. 

\vspace{1em}
Unlike the Cartesian lattice, there seems to be no systematic way of coding hexagonal images. This is the void that our work attempts to address. In our work, the adaptive tree-based sub-band coding techniques are explored for hexagonal image processing. The reason for our choice is that DWT-based methods are general and usually outperform DCT-based methods~\cite{ouafi2008modified}. Subsequently, they produce fewer compression artifacts at lower bits-per-pixel (BPP) rates compared to other transform-based methods. 
After reviewing related DWT-based compression schemes for hexagonal images, we are affirmed that there is scope for significant improvement in this field. Previous works lack the application of standard coding schemes, hexagonal sampling methods and associated multiresolution analysis techniques. Therefore, we attempt to bridge this gap in hexagonal image compression by concentrating on sub-band coding systems.

\section{Background}
\label{sec:background}

\subsection{2-D Index Mapping}
\label{sec:2-D_index_mapping}
The hexagonal lattice is generated by the matrix 
\begin{equation} \label{eq:basis_hex}
\mathbf{V} := \bordermatrix{~ & \mathbf{v_1} & \mathbf{v_2} \cr
                  ~ & 1 & -\rfrac{1}{2} \cr
                  ~ & 0 & \rfrac{\sqrt{3}}{2} \cr}.
\end{equation}
Given a scaling factor (or sampling interval) $h$, any point on the (scaled) lattice is produced by the matrix-vector product $h\mathbf{V}{\mathbf{k}}$ where $\mathbf{k} \in \mathbb{Z}^2$ is a 2-D integer vector that defines an integer coordinate system on the lattice. Given a bivariate function $f(\mathbf{x})$ ($\mathbf{x} \in \mathbb{R}^2$), a sampled version of the function on the hexagonal lattice is given by the sequence
\begin{equation} \label{eq:sampling_eqn}
F[\mathbf{k}] := \{ f(h \mathbf{V}{\mathbf{k}} ), {\mathbf{k}} \in \mathbb{Z}^2 \}.
\end{equation}

Since $\mathbf{k} \in \mathbb{Z}^2$, a 2-D index can be used to map the hexagonal lattice into a rectangular memory region (2-D index map). The main advantage of a rectangular index map is that it lends itself to a straightforward adoption of rectangular convolution routines already implemented for the square lattice. In wavelet transformation, convolution plays a pivotal role. It is crucial to have efficient access to the neighbors of a pixel. With a rectangular index map, the neighbors can be fetched in constant time via simple indexing operations. 
If the convolution filters (as shown in Fig.~\ref{fig:dhwtfilters}) are converted accordingly into index maps, then convolution on the hexagonal lattice is executed in accustomed runtime.

\begin{figure}[!h]
	\centering
	\includegraphics[width=\columnwidth]{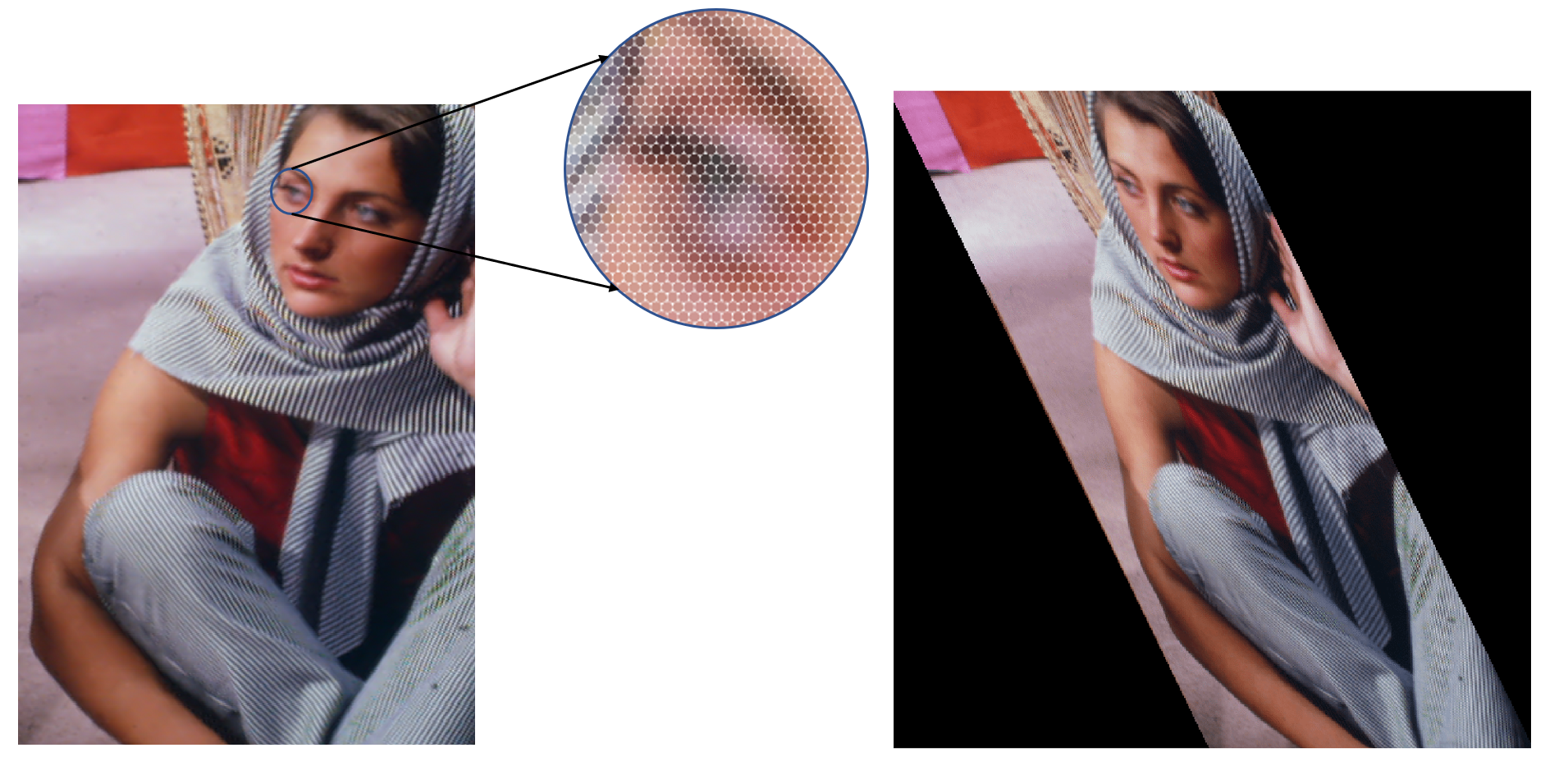}
    \begin{minipage}[c]{.4\linewidth}
    
    \captionsetup{singlelinecheck=false, justification=raggedright}%
    \subcaption{Hexagonal lattice}\label{hex_lattice_barbara}
    \end{minipage}%
    \captionsetup{singlelinecheck = false, justification=raggedleft}
    \begin{minipage}[c]{.5\linewidth}
    \subcaption{2-D index mapping}\label{2d_index_mapping_barbara}
    \end{minipage}
     \captionsetup{singlelinecheck = true, justification=raggedright}%
	\caption[2-D index mapping of Barbara image]{(b) 2-D memory array (index map of size $511 \times 511$) is used to store the (a) hexagonal lattice of the Barbara image (of size $256 \times 256 \times 2$). Black colored cells represent the extra padded zero values in the 2-D index map.}
	\vspace{-1em}
	\label{fig:indexmap}
\end{figure}

Images are typically captured via rectangular sensors. In order to represent a rectangular spatial region on a hexagonal lattice via a 2D index table, we use a parallelogram in the table with triangular zero padded regions appended as shown in Fig.~\ref{fig:indexmap}. The trade-off is extra memory which is required to store a rectangular image using such a coordinate system. The number of rows in the index map is the same as the number of rows of the rectangular region of the hexagonal lattice. However, the number of columns is twice the number of pixels in the columns of the hexagonal grid.  The succession and placement of the rows are set according to the arrangement of the corresponding pixel positions in the hexagonal grid. 

After producing the index map from the hexagonal lattice, the forward wavelet transform is performed to capture the change of frequency and location information. Then the encoding process commences to reorder the wavelet coefficients based on their significance in such a way that they can be compressed efficiently. 

\subsection{Discrete hexagonal wavelet transform}
\label{sec:multiresolution}

A discrete wavelet transform on the hexagonal lattice is similar to the Cartesian lattice provided that analysis and synthesis filters that respect the hexagonal lattice are employed. In our work, we employ the compactly supported bi-orthogonal filters proposed by Cohen and Shlenker~\cite{Cohen1993}. The important steps of the transformation process are summarized as follows.

For a given bivariate function $f(\mathbf{x})$ that is sampled on a hexagonal lattice, a linear approximation from samples $F[\mathbf{k}]$ is given by
 \begin{equation} \label{eq:linear_approximation}
 f(\mathbf{x}) \approx \tilde{f}(\mathbf{x}) = \sum_{\mathbf{k} \in \mathbb{Z}^2} F[\mathbf{k}] \Phi (\mathbf{x - Vk}),
 \end{equation}
 where $\Phi(\mathbf{x}) = \Phi(x_1, x_2)$ is a scaling function that satisfies the following dyadic scaling relationship.
  \begin{equation} \label{eq:wavelet_dec2}
 \Phi(\mathbf{x}) = \sum_{\mathbf{k}} w_0[\mathbf{k}] \Phi (2\mathbf{x - Vk}). 
 \end{equation}
 The corresponding wavelet functions $\Psi_i(\mathbf{x})$ ($i \in \{ 1, 2, 3\}$) also satisfy a similar dyadic scaling relationship.
 \begin{equation} \label{eq:wavelet_dec3}
 \Psi_i(\mathbf{x}) = \sum_{\mathbf{k}} w_i[\mathbf{k}] \Phi (2\mathbf{x - Vk}), \ i \in \{ 1, 2, 3\}.
 \end{equation}
 In Equations~\ref{eq:wavelet_dec2} and~\ref{eq:wavelet_dec3}, $w_0$ and $w_i$ constitute a set of \emph{synthesis} filters which hold the coefficients of the scaling and wavelet functions respectively. Denoting the coarse approximation as $C[\cdot]$ and detail coefficients as $D_i[\cdot]$, the coarse-to-fine reconstruction process can be written as
 \begin{equation} \label{eq:dwt}
  \tilde{f}(\mathbf{x})= \sum_{\mathbf{k}} C[\mathbf{k}] \Phi (\tfrac{\mathbf{x}}{2} - \mathbf{Vk}) \ + \sum_{i = 1}^{3} \sum_{\mathbf{k}} D_i[\mathbf{k}] \Psi_i (\tfrac{\mathbf{x}}{2} - \mathbf{Vk}),
 \end{equation}
 where the coarse approximations $C$ and details $D_i$ are obtained by convolving $F$ with appropriate \emph{analysis} filters $\mathring{\omega}_i$ where $ i \in \{0, 1, 2, 3\}$. Specifically,
\begin{equation} \label{eq:wavelet_dec1}	
C[\mathbf{k}] = (F \ast \mathring{\omega}_0) [2\mathbf{k}], \ \\
D_i[\mathbf{k}] = (F \ast \mathring{\omega}_i) [2\mathbf{k}] \ \text{for} \ i \in \{1,2,3\},
\end{equation}
where $\mathring{\omega}_0$ is a low-pass analysis filter and $\mathring{\omega}_i, ( i \in \{1, 2, 3\})$ are three high-pass analysis filters. 

\begin{figure}[t]
	\centering
	\includegraphics[width=\columnwidth]{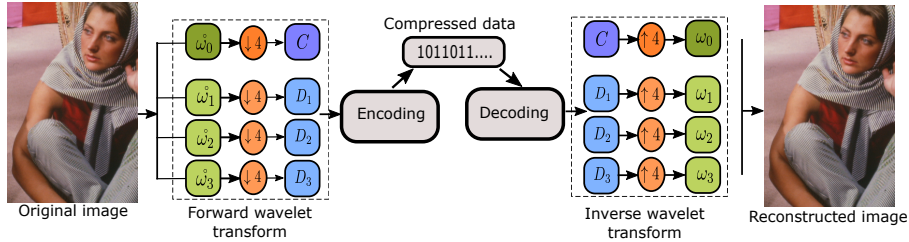}
	\caption{Sub-band decomposition-reconstruction process. A four channel filter bank separates the source input into frequency bands through filtering and downsampling. Then encoding-decoding is performed. Finally, the decoded output is reassembled via upsampling and filtering followed by an inverse discrete wavelet transform.}
	\label{fig:sub-band_coding_system}
	\vspace{-1em}
\end{figure}

Let $C'$ and $D_i'$ denote dyadically upsampled versions of $C$ and $D_i$ respectively. The reconstruction process consists of 2-D discrete convolutions with the synthesis filters and can be written as follows.
\begin{equation} \label{eq:wavelet_rec1}
F[\mathbf{k}] = (C' \ast \omega_0) [2\mathbf{k}] + \sum_{i=1}^3 (D'_i \ast \omega_i) [2\mathbf{k}].
\end{equation}
Fig.~\ref{fig:sub-band_coding_system} represents this transformation process schematically.

\begin{figure}[htbp]
	\centering
	\includegraphics[width=\columnwidth]{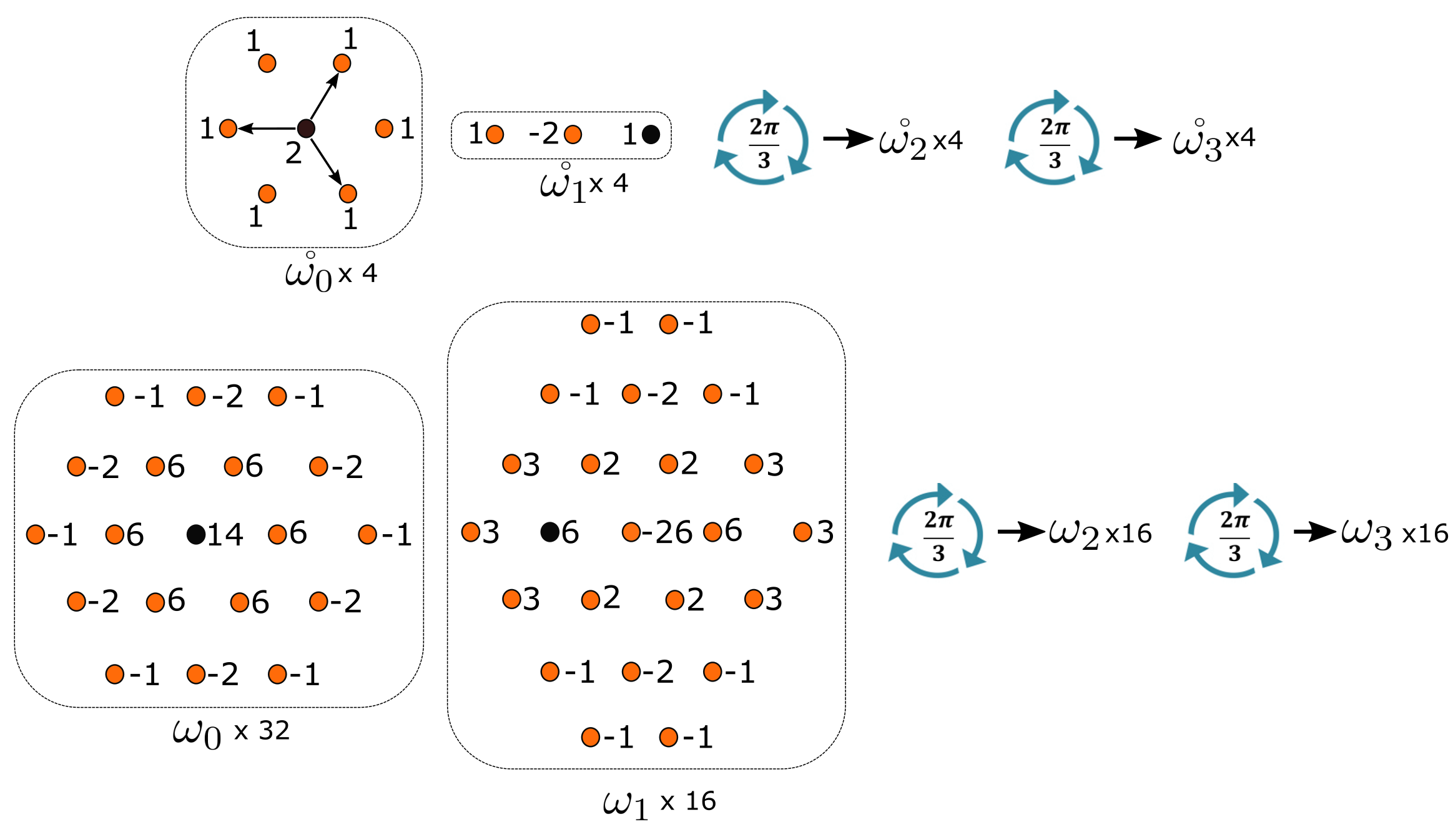}
	\caption[Scale co-efficients of DHWT]{Scale co-efficients of decomposition filters $\mathring{\omega}_0, \mathring{\omega}_1, \mathring{\omega}_2, \mathring{\omega}_3$ and reconstruction filters ${\omega}_0, {\omega}_1, {\omega}_2, {\omega}_3$ for DHWT. The decomposition filters $\mathring{\omega}_2$ and $\mathring{\omega}_3$ are obtained by successive $\frac{2\pi}{3}$ rotations of $\mathring{\omega}_1$. Similarly, reconstruction filters ${\omega}_2$ and ${\omega}_3$ are obtained by successive $\frac{2\pi}{3}$ rotations of ${\omega}_1$. All the filters are 2-D and non-separable. Black colored points indicate the position of the origin.}
	\label{fig:dhwtfilters}
	\vspace{-1em}
\end{figure}

The discrete hexagonal wavelet transform (DHWT) is a non-separable scheme designed for the hexagonal lattice. The biorthogonal hexagonal wavelet bases proposed by Cohen and Schlenker~\cite{Cohen1993} are compactly supported and ensure perfect reconstruction using the decomposition and reconstruction filters shown in Fig.~\ref{fig:dhwtfilters}. The low pass analysis filter $\mathring{\omega}_0$ yields the linear spline on the hexagonal lattice which is also known as the Courant interpolating function~\cite{Cohen1993}. The linear spline is a box spline that provides a second-order approximation on the hexagonal lattice~\cite{entezari08}. Hence, the number of vanishing moments for the corresponding wavelets is 2. 
The high-pass analysis filters $\mathring{\omega}_i, ( i \in \{1, 2, 3\})$ of DHWT compute directional derivatives in three principal directions of the hexagonal lattice. 
The construction procedure of Cohen and Schlenker~\cite{Cohen1993} can be used to produce higher regularity wavelets on the hexagonal lattice. In this work however, we restrict attention to the second-order wavelet as our focus is on the coding of the wavelet coefficients. It is important to note that our proposed coding scheme is not restricted to the second-order wavelet and higher order wavelets can also be used. This is a subject of future investigation.

\begin{figure}[!h]
	\centering
	\includegraphics[scale=0.3]{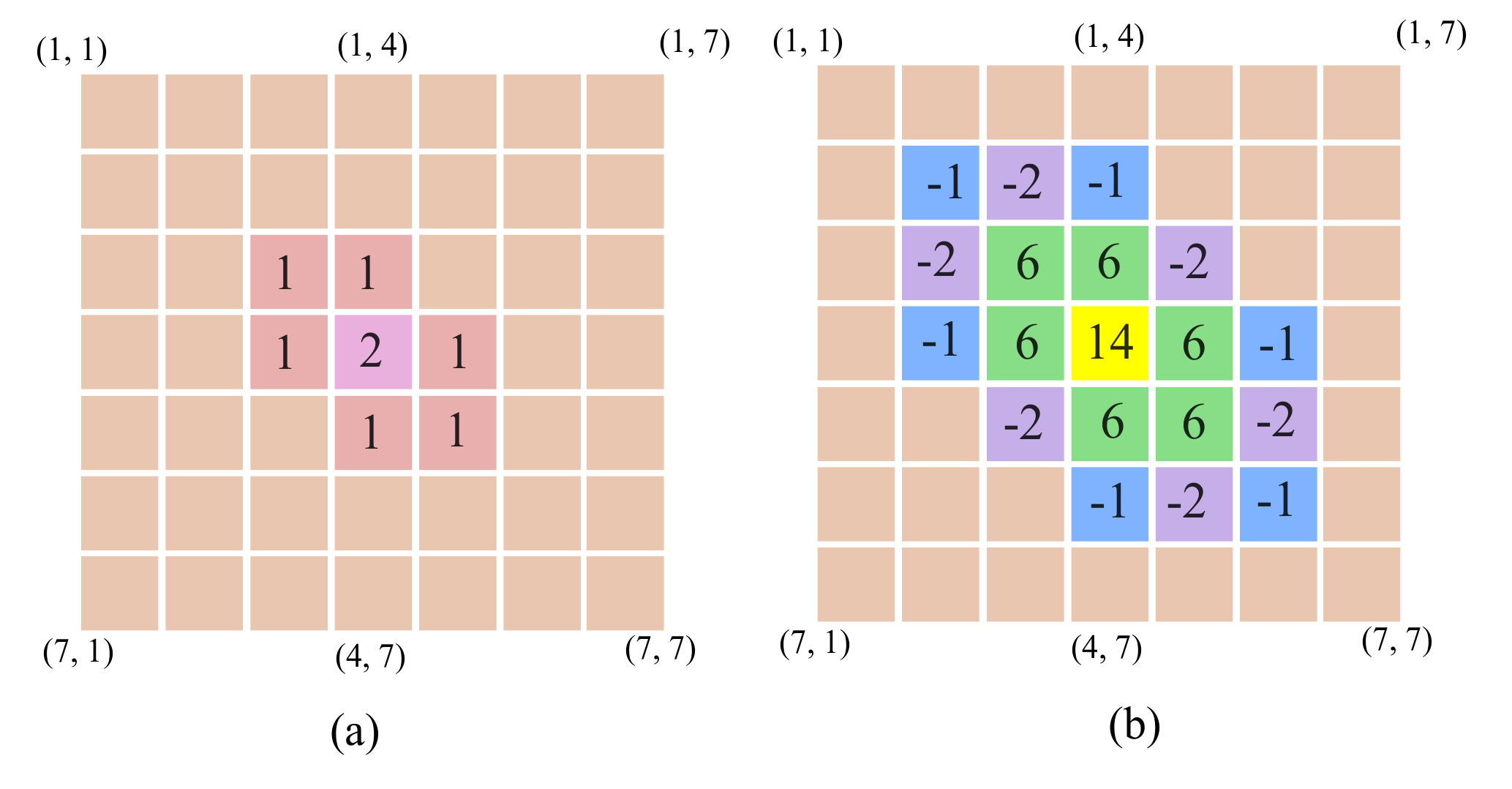}
	\caption[2-D index mapping of DHWT filters]{2-D index map (of size $7 \times 7$) of (a) decomposition low-pass filter  $\mathring{\omega}_0$ and (b) reconstruction filter ${\omega}_0$. Actual hexagonal structures of $\mathring{\omega}_0$ and ${\omega}_0$ are shown in Fig~\ref{fig:dhwtfilters}. In the figure, the indices are integer values, and range from 1 to the length of the row or column.}
	\label{fig:indexed_map_filters}
	\vspace{-1em}
\end{figure}

After producing the index map from the hexagonal lattice, the forward wavelet transform is performed to capture the directional information. To correlate with the index map of the source image, the DHWT filters are transformed into equivalent 2-D index maps. Each filter is converted into a $7 \times 7$ 2-D matrix. The origin of the filters is set at the center of the index map. The non-separability of the filters is preserved during the mapping from the original lattice. Two examples of filters stored in the index maps are shown in Fig.~\ref{fig:indexed_map_filters}.

\begin{figure}[!h]
	\centering
	\includegraphics[scale=0.2]{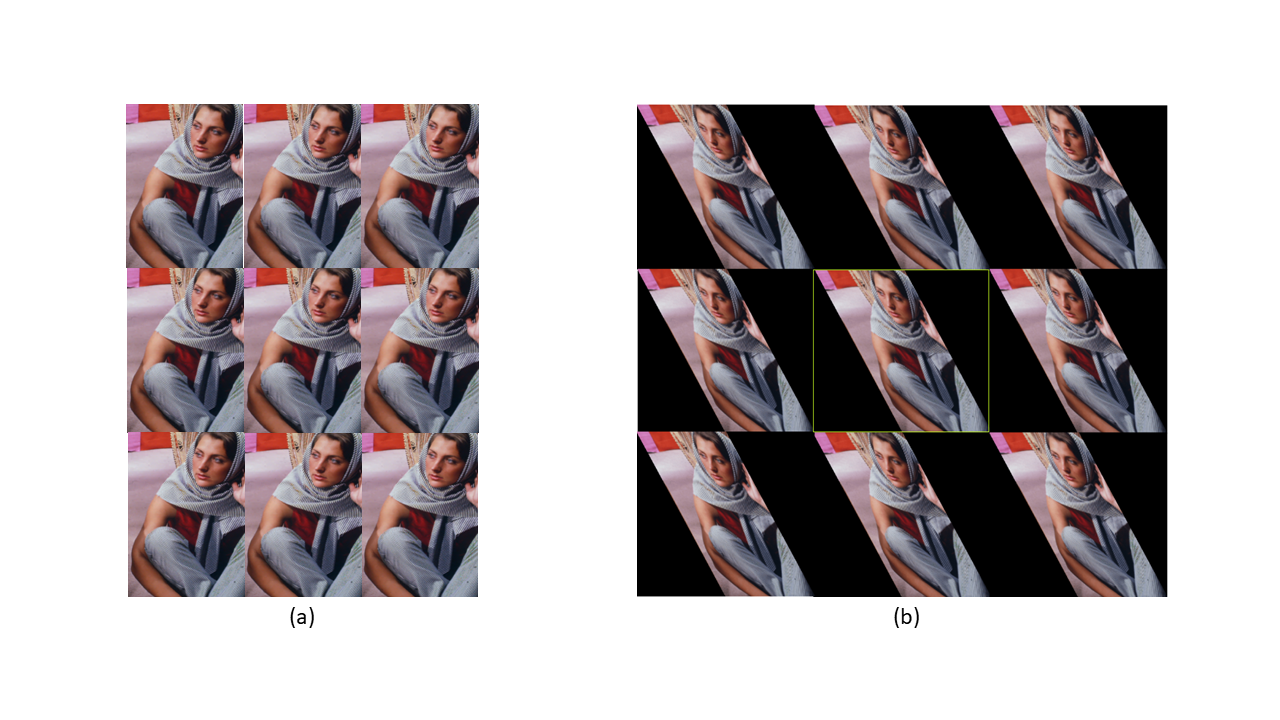}
	\caption[Periodic extension on 2-D grids]{Periodic extension of 2-D index map for the Barbara image on (a) Cartesian grid, and (b) hexagonal grid. The extension is rectangular where the source 2-D index map (center) is surrounded by copies of the source as shown.}
	\vspace{-1em}
	\label{fig:periodization}
\end{figure}

As the filters (illustrated in Fig.~\ref{fig:indexed_map_filters}) have wide support, border distortion is expected during the wavelet transform. Consistent boundary conditions are required to circumvent boundary effects~\cite{mallat1999wavelet, deQueiro1992Spof}. In our tests, extra padding in the boundaries is done according to a periodized extension~\cite{strang:nguyen:2009} for both Cartesian and hexagonal images. The periodic extension introduces edge discontinuities for both the Cartesian and hexagonal grids as shown in Fig.~\ref{fig:periodization}. 
For a 1-D signal $ x_1 \ x_2 \ ... \ x_n $, the periodized extension adjoining the boundaries is as follows: $... x_{n-1} \ x_n \ | \ x_1 \ x_2 \ ... \ x_n \ | \ x_1 \ x_2 \  ... \ $.
This mode of extension assumes that the signal or image is repeated at the boundaries. If the signal length is odd, the signal is first extended by adding an extra sample equal to the last value on the right. Then a periodized extension is performed on each side. While more sophisticated boundary conditions could be used, we stick to this periodized extension as it is simple to implement and also provides perfect reconstruction with the wavelet filters shown in Fig.~\ref{fig:dhwtfilters}.

\section{SBHex}
\label{sec:SBHex}

\subsection{Overview}
\label{sec:SBHexOverview}
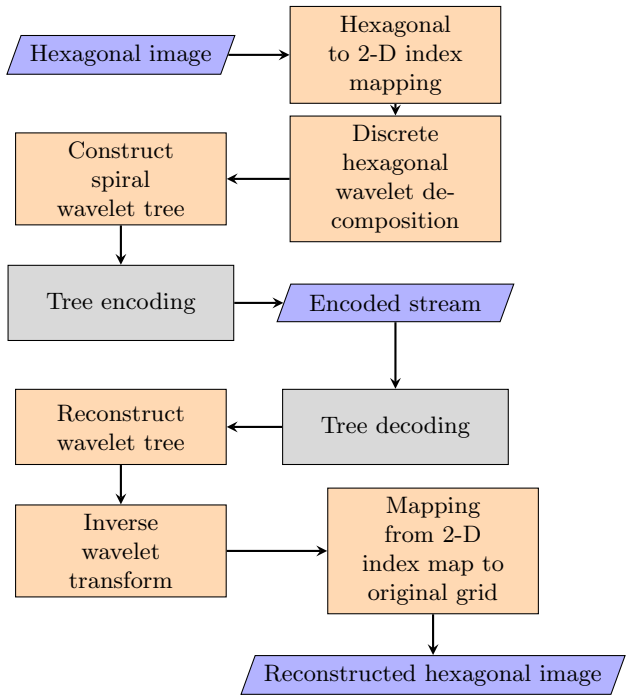
\begin{figure}[!tbp]
\centering
    \begin{tikzpicture}[node distance=1.65cm]
        \tikzstyle{every node}=[font=\small]
            \node (in1) [io] {Hexagonal image};
            \node (pro1) [process, right of=in1, xshift=2cm] {Hexagonal to 2-D index mapping};
            \node (pro2) [process, below of=pro1] {Discrete hexagonal wavelet decomposition};
            \node (pro3) [process, left of=pro2, xshift=-2cm] {Construct spiral wavelet tree};
            \node (pro4) [process2, below of=pro3] {Tree encoding};
            \node (in2) [io, right of=pro4, xshift=2cm] {Encoded stream};
            \node (pro5) [process2, below of=in2] {Tree decoding};
            \node (pro6) [process, left of=pro5, xshift=-2cm] {Reconstruct wavelet tree};
            \node (pro7) [process, below of=pro6] {Inverse wavelet transform};
            \node (pro8) [process, right of=pro7, xshift=2.5cm] {Mapping from 2-D index map to original grid};
            \node (out) [io, below of=pro8] {Reconstructed hexagonal image};

        \draw [arrow] (in1) -- (pro1);
        \draw [arrow] (pro1) -- (pro2);
        \draw [arrow] (pro2) -- (pro3);
        \draw [arrow] (pro3) -- (pro4);
        \draw [arrow] (pro4) -- (in2);
        \draw [arrow] (in2) -- (pro5);
        \draw [arrow] (pro5) -- (pro6);
        \draw [arrow] (pro6) -- (pro7);
        \draw [arrow] (pro7) -- (pro8);
        \draw [arrow] (pro8) -- (out);
    \end{tikzpicture}
    \caption{Pipeline of SBHex sub-band coding} 
    \label{fig:sbhex_pipeline}
    \vspace{-1em}
\end{figure}

Our focus in this work is on coding/decoding the sub-band coefficients $C$ and $D_i$ (Fig.~\ref{fig:sub-band_coding_system}).
Our first proposed method for hexagonal sub-band coding is SBHex (Spirally-mapped Branch-coding for Hexagonal images). The main difference between SBHex and conventional tree-based coding schemes is that SBHex uses a different data structure (or wavelet orientation tree). It utilizes the discrete hexagonal wavelet transformation (DHWT) and the isotropic nature of the distribution of significant pixel coefficients in the low-frequency components. An outline of the encoding-decoding pipeline of SBHex is shown in Fig.~\ref{fig:sbhex_pipeline}. 

\begin{figure}[!tbp]
    \centering
    \includegraphics[scale=0.25]{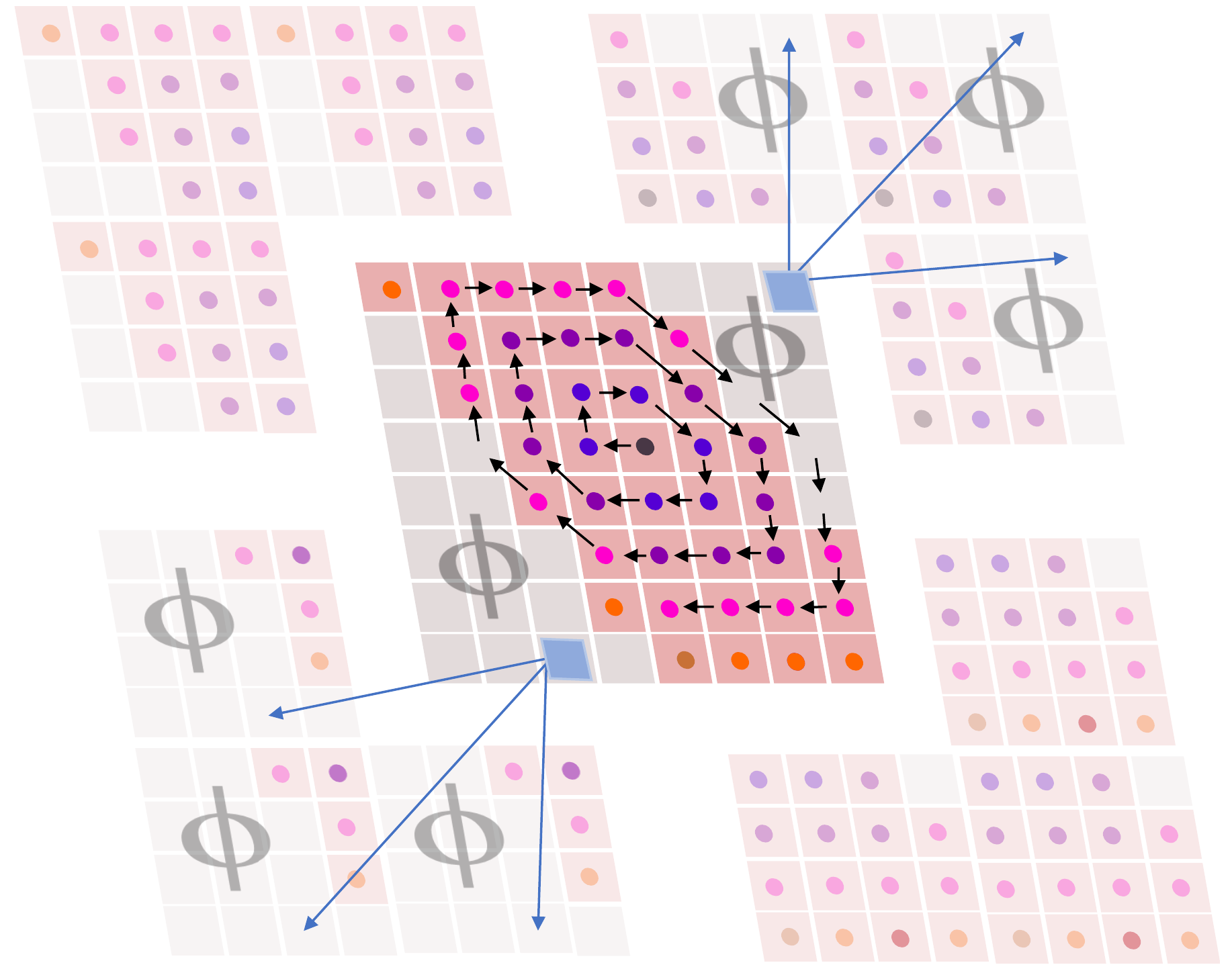}
    \caption{In SBHex, the detail coefficients surround the coarse coefficients. Our hexagonally spiral coding scheme respects the orientation of the DHWT produced sub-bands. It captures the most significant coefficients earlier and locates the padded zeroes at the end of the encoded stream.} 
    \label{fig:sbhex_mapping}
    \vspace{-1.0em}
\end{figure}

SBHex commences with 2-D index mapping from a hexagonal lattice (described in Section~\ref{sec:2-D_index_mapping}). After that, a forward DHWT is performed in the preferred multiresolution level. At this point, we get coarse scale coefficients and a set of details. Using the different set of sub-bands, a spiral tree is created whereby the coarse scale is present at the top level and the detail blocks (or branches) are spirally ordered according to their location in the sub-bands as depicted in ~Fig.~\ref{fig:sbhex_mapping}. The encoding and quantization are performed from the center of the coarse scale by traversing in a spiral fashion as described below. 

\subsection{Spiral mapping}
The DHWT captures directional variations of an image and results in wavelet coefficients. After performing several levels of hierarchical decomposition, there are wavelet coefficients in different sub-bands that represent the same spatial location in the hexagonal image. Natural images are decomposed in a way that most energy is compacted into lower bands. In our proposed spiral mapping, we organize the coefficients in a tree structure such that higher magnitude coefficients have a high probability of being close to the root of the tree compared to coefficients with lower magnitudes. 

\begin{figure}[tbp!]
	\centering
	\includegraphics[scale=0.25]{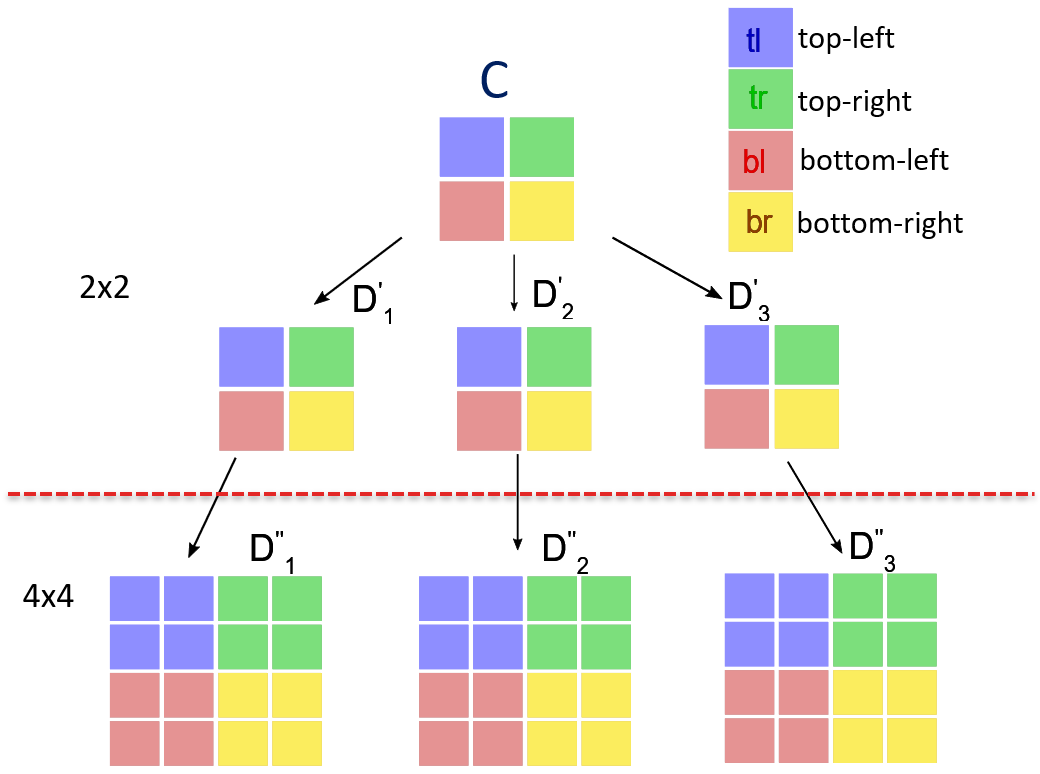}
	\caption{(a) Wavelet sub-tree structure of a $16 \times 16$ index map where the first level of detail $D''$ consists of $4 \times 4$ blocks and the second level of detail $D'$ consists of $2 \times 2$ blocks. Finally, the lowest level (or coarse band) is a $2 \times 2$ block. Each of the sub-bands is split into four equal sized quadrants and arranged according to the position and level of decomposition.}
	\label{fig:band_splitting}
	\vspace{-1em}
\end{figure}

In order to ensure that insignificant coefficients are kept together in large subsets, before we begin the mapping process, we also split each of the bands into four quadrants as shown in Fig.~\ref{fig:band_splitting}.
Each of the sub-bands is split into four equal regions as follows:
\begin{itemize}
 	\item $tl$ - Top-Left region of the sub-band
 	\item  $tr$ - Top-Right region of the sub-band
 	\item  $bl$ - Bottom-Left region of the sub-band
 	\item  $br$ - Bottom-Right region of the sub-band
\end{itemize}
  
\begin{figure}[htbp]
	\centering
	\includegraphics[scale=0.25]{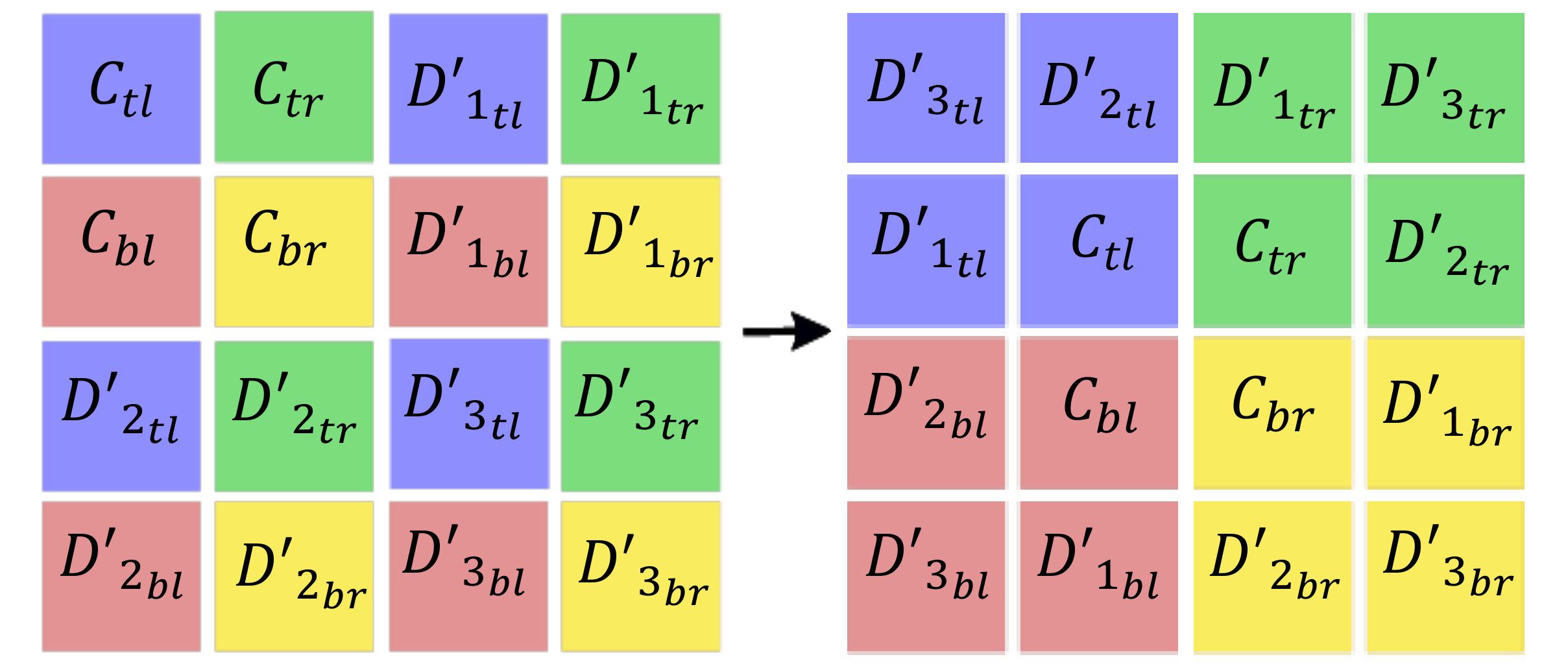}
	\caption{Pictorial comparison of conventional hierarchical mapping (left) vs. spiral mapping (right) for a $4 \times 4$ image. The multiresolution analysis is performed to decomposition level 1. Coarse scale block $C$, and detail blocks $D'_1$, $D'_2$ and $D'_3$ are each divided into quadrisections (left) and rearranged as shown (right).}
	\label{fig:spiral_mapping}

\end{figure}

\begin{figure}[!tbp]
\centering
  \begin{subfigure}[b]{3.5cm}
  \centering
    \includegraphics[scale=0.5]{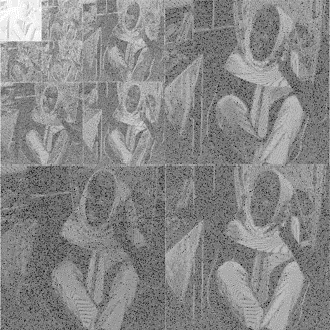}
    \caption{Hierarchical mapping}
  \end{subfigure}
  \hspace{1cm}
  \begin{subfigure}[b]{4cm}
  \centering
    \includegraphics[scale=0.5]{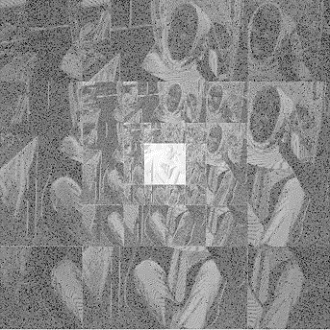}
    \caption{Spiral mapping}
  \end{subfigure}
  \caption{Pictorial comparison of (a) hierarchical mapping and (b) spiral mapping of wavelet coefficients of Barbara image (based on Fig.~\ref{fig:spiral_mapping}). The multiresolution analysis is performed to decomposition level 3. For ease of comparison, Cartesian wavelet coefficients are shown.}
  \label{fig:spiral_mapping_barbara}
  \vspace{-1em}
\end{figure}

The blocks of coefficients are then rearranged into a spiral spatial orientation tree as shown in Fig.~\ref{fig:spiral_mapping}. 
The coarse scale coefficients $C$ are placed in the center and corresponding quadrants of $D_1$, $D_2$, and $D_3$ are arranged in a clock-wise manner around the coarse scale coefficients (Fig.~\ref{fig:spiral_mapping}). This pattern is then repeated for additional levels of detail (Fig.~\ref{fig:spiral_mapping_barbara}).
This organization aggregates a contiguous group of blocks that are associated with distinct frequency bands. If we scan the coefficients from the center of the coarse scale and traverse spirally, we will pass from lower to higher levels in a manner that gives equal importance to all detail bands rather than visiting one band completely before the other as is the case in Morton scan. The precise scanning pattern we use is described in Sec.~\ref{sec:hexScanning}.

This configuration of the spatial orientation tree also benefits the encoding of the extra padded zeros which are placed at the end of the pass. This results in the concatenation of trailing zero trees and isolated zeros at the end of significance codestream. The compression is unaffected as it discards the trailing insignificant symbols since they do not transmit any information while decoding.

\subsection{Parent-to-children relationship}
\label{sec:parent2childrenreln}
All of the descendants of the parent coefficient are referred to as children of that coefficient. These include immediate children and grandchildren and so on. The parent-to-children relationship is configured based on the position of spirally arranged blocks of sub-bands. Moving from the center to the surroundings, each block in the next lower level in each band has twice as many rows and columns as the block in the level above it. Thus, each block at the next lower level has four times as many coefficients as the block in the level above it. 
Let us assume that a 2-D index map of dimension $R \times C$ is spirally mapped into a spiral wavelet tree (of the same dimension) where $R$ and $C$ are even integers by definition and represent the number of rows and columns respectively. Let $(r,c)$ denote the coordinates of an entry in the spiral wavelet tree. Using a one-based row-column indexing scheme, these coordinates range between $(1, 1)$ (upper-left) and and $(R, C)$ (bottom-right). 

\subsubsection{Roots of Quadrisection}
The spiral wavelet tree is split into quadrisection. Each of the quadrisection has its root sourced at the inner core of the wavelet tree. 
The top-left quadrisection has the parent-root at index $(\frac{R}{2}, \frac{C}{2})$, the top-right quadrisection has its parent-root at index $(\frac{R}{2}, \frac{C}{2} + 1)$, bottom-left has its parent-root at index $(\frac{R}{2} + 1, \frac{C}{2})$, and finally the bottom-right quadrisection has its parent-root at index $(\frac{R}{2} + 1, \frac{C}{2} + 1)$. 

\subsubsection{Children}
\begin{figure}[tbp!]
	\centering
    \includegraphics[scale=0.3]{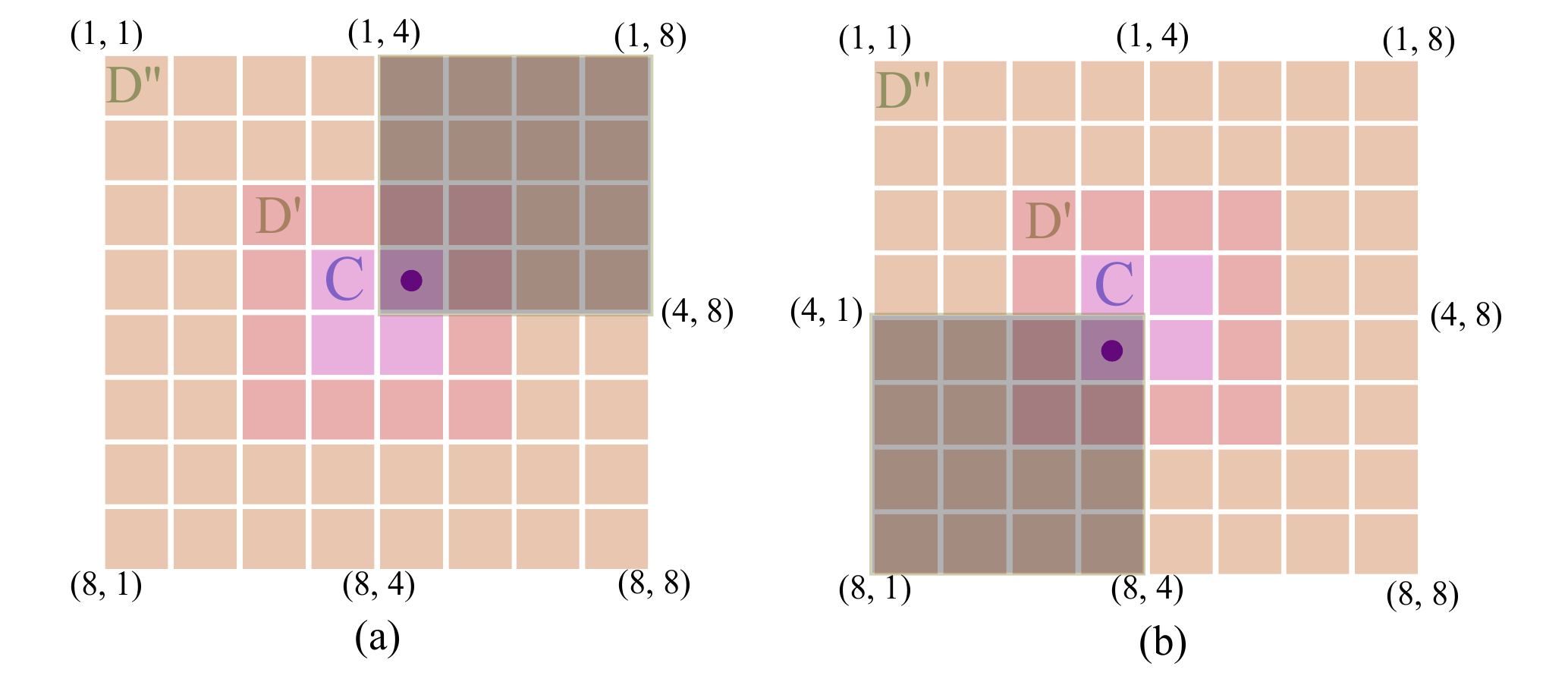}
	\caption{In an $8 \times 8$ spiral wavelet tree, (a) for parent $(4, 5)$, all the coefficients at the top-right are the children. (b) If the parent is $(5, 4)$, then all the coefficients at the bottom-left are its children. In the wavelet tree, $C$ means coarse approximation. $D'$ and $D''$ are the set of details. The multiresolution analysis is performed to level 2.}
	\label{fig:p2c_mask_root}
	\vspace{-1.0em}
\end{figure}

\begin{figure}[!tbp]
	\centering
	 \includegraphics[scale=0.3]{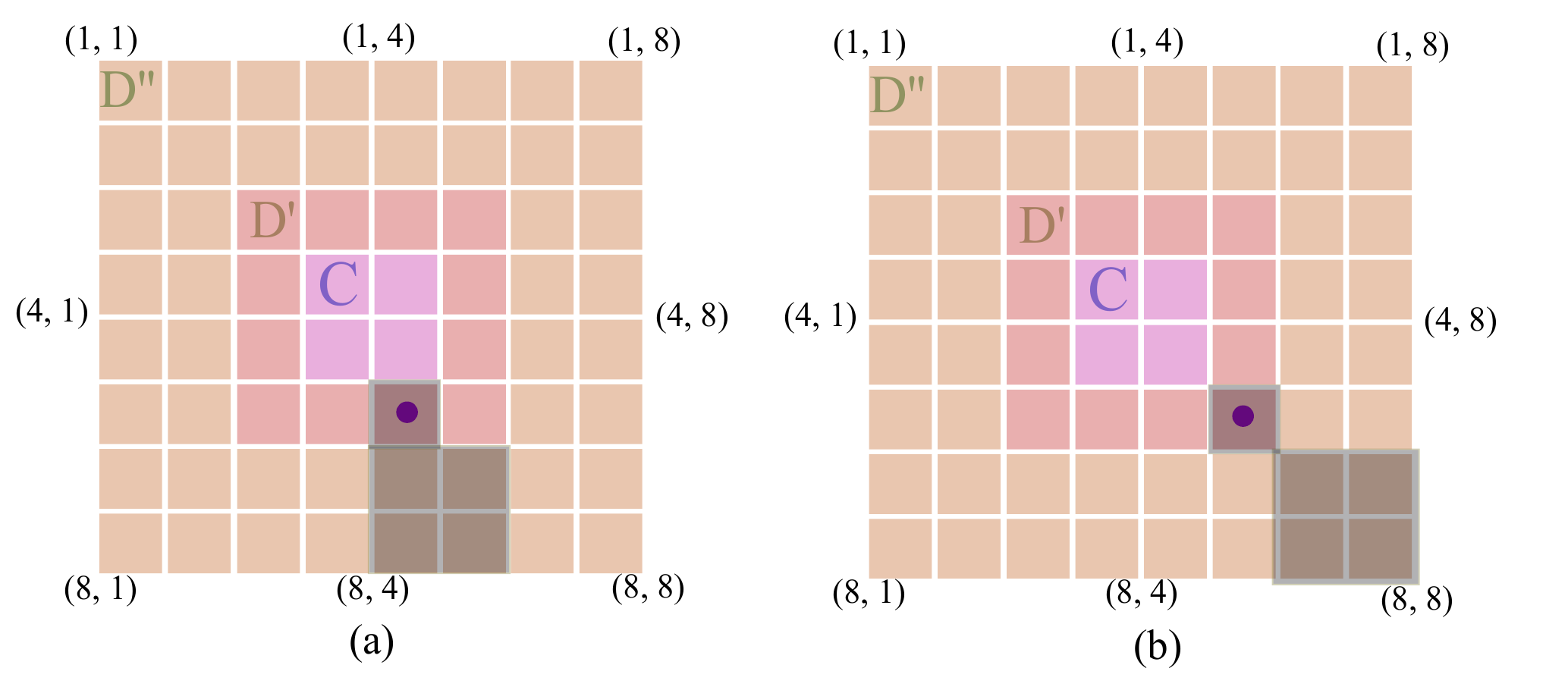}
	\caption{In an $8 \times 8$ spiral wavelet tree, (a) for parents $(6, 5)$ and (b) $(6, 6)$, the immediate four children at the next level are shown. Here, $C$ refers to coarse approximation coefficicents while $D'$ and $D''$ are detail coefficients. The multiresolution analysis is performed to decomposition level 2.
	}
	\label{fig:p2c_8x8_1}
	\vspace{-1.0em}
\end{figure}

In order to capture most of the coefficients from a root-parent as an insignificant tree, all the values in a particular direction are considered as children of the root-parent as shown in Fig.~\ref{fig:p2c_mask_root}. 
Except the root of quadrisections of the wavelet tree, each coefficient in a given block has four children in corresponding locations in the next level. The four immediate children of the coefficient at index $(r, c)$ are located at $\left(2r - (R/2 + 1), 2c - (C/2 + 1)\right)$ (top-left child), $\left(2r - (R/2 + 1), 2c - C/2\right)$ (top-right child), $\left(2r - R/2, 2c - (C/2 + 1)\right)$ (bottom-left child) and $\left(2r - R/2, 2c - C/2\right)$ (bottom-right child).
Two examples of this relationship are shown in Fig.~\ref{fig:p2c_8x8_1}. 

\subsection{Encoding-Decoding}

Our encoder starts with a magnitude test of the coefficients in order to find the `significant' coefficients: \emph{significant-positive} (p) and \emph{significant-negative} (n), compared to a \emph{threshold}. Considering $C_{max}$ as the most significant coefficient in the wavelet tree structure, the initial threshold is set to $2^{\lfloor \log_2 C_{max} \rfloor}$. After each iteration (in successive passes), the threshold is reduced to half of the previous magnitude. For encoding, the \emph{dominant pass} (EZW) is executed based on user-preference to generate the compressed code stream from the spiral wavelet tree. In case of the dominant pass, an arithmetic coding of the symbol stream is required to compress more information. So, the significance map produced by the dominant pass is coded further using Huffman coding according to Table~\ref{table:huffman_coding}. Considering the possibility of a higher number of \emph{zero-trees} (t) and \emph{isolated zeroes} (z), their corresponding symbols are assigned smaller codes.

\begin{table}[!ht]
\centering
\begin{tabular}{|c|l|l|}
\hline
Symbol & Meaning & Code\\\hline
t & zero-tree & 0 \\
z & isolated zero & 10 \\
n & significant-negative & 110 \\
p & significant-positive  & 1110 \\
separator & splits each dominant pass& 1111\\
\hline
\end{tabular}
\caption{Huffman codes for symbols} 
\label{table:huffman_coding}
\vspace{-1.0em}
\end{table} 

Finally, a \emph{refinement pass} is performed for quantizing the significant values into the subordinate list. \emph{Refinement pass} uses a scalar uniform quantizer which reduces the precision of values to some fixed number of levels across the range of uniformly spaced quantization levels.

Decoding the tree from the compressed code is obtained by reversing the encoding process. At the beginning of the decoding process, an empty wavelet transform array (the same size as the original wavelet coefficient tree) is initialized. The wavelet tree is filled up by decoding of the significance map. At the end of the decoding process, the wavelet tree is regenerated based on the BPP setting preferred by user. At this point, the reconstructed image is obtained by applying the inverse wavelet transform and remapping the blocks to the index map of the image.

\subsection{Hexagonal scanning order}
\label{sec:hexScanning}
For traversing in the spiral wavelet tree, we implement a hexagonal scanning order which respects the underlying hexagonal grid in the index map. The scan order starts from the center of the grid. The sequence of change of direction repeatedly occurs in a clock-wise manner.
 
\begin{figure}[ht]
    \centering
    \includegraphics[scale=0.2]{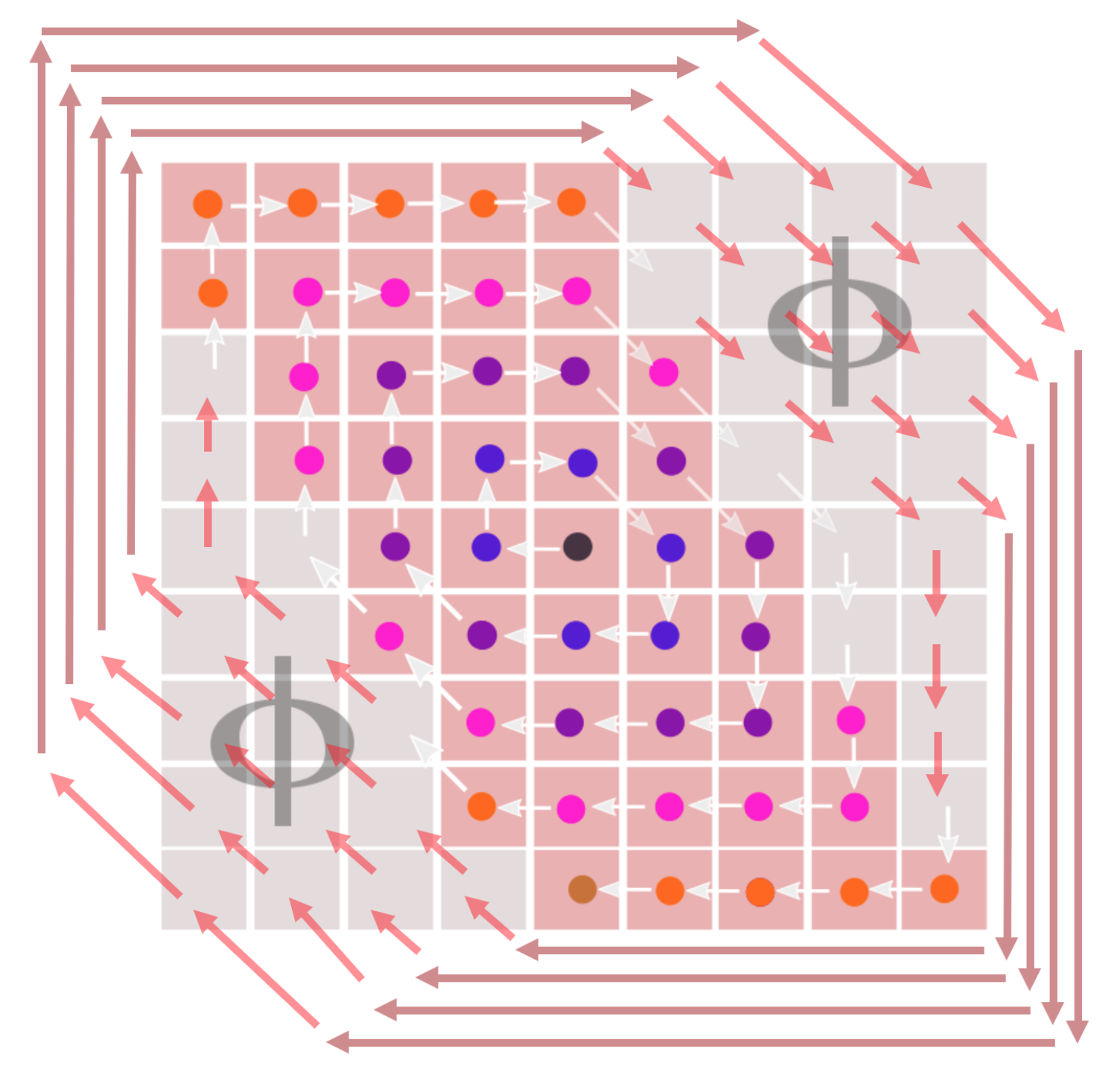}
    \caption{Hexagonal traversal of spirally mapped wavelet tree. Only the coarse scale is shown for simplicity.}
    \label{fig:hex_planefilling_curve}
    \vspace{-1.0em}
\end{figure}

For each cycle, away from the center, the number of steps that point in the same direction increments in the scan order. For example, in Fig.~\ref{fig:hex_planefilling_curve}, we can observe that as the row number away from the center increases, the number of steps that go from left to the right increases. The cyclic order of the scan involves scanning through the coordinates considering the underlying hexagonal geometry of the wavelet. The purpose is to scan the coefficients from the center and diverge out to the boundary region as the index map is girded by extra padded zeros at the boundaries. Moreover, subjects of interest are usually placed at the center of in photographic images. To scan all the coefficients in the spiral wavelet tree, the hexagonal scan order is required to be a space-filling curve. A space-filling curve is a curve whose traversal ensures passing through all the points on the given 2-D grid~\cite{valgaerts2005space}. Our hexagonal scanning order is also capable of generating space-filling curves for variable-sized 2D grids. An illustration on a $9 \times 9$ array is shown in Fig.~\ref{fig:hex_planefilling_curve}. For a 2D rectangular array, this scanning pattern will eventually go through cells that are outside the array. With a simple boundary check, cells outside the array are skipped until the scanning path renters the array as shown in Fig.~\ref{fig:hex_planefilling_curve}. From the figure,  we can observe that the extra padded zeros are scanned later. During the encoding process, these extra padded zeros generate trailing zero-trees or isolated zeros in the codestream which do not contribute in the reconstruction (during decoding). Therefore, the concatenated zero-trees or isolated zeros are discarded from the codestream and do not affect the space-saving capability of the encoded information. Furthermore, when applied to a spiral wavelet tree, this scanning order cycles through the detail subbands giving equal priority to all three subbands owing to the spiral arrangement of the bands (see Fig.~\ref{fig:spiral_mapping} and Fig.~\ref{fig:spiral_mapping_barbara}).

\section{BBHex}
\label{sec:bbhex}
\begin{figure}[!t]
	\centering
	\includegraphics[scale=0.3]{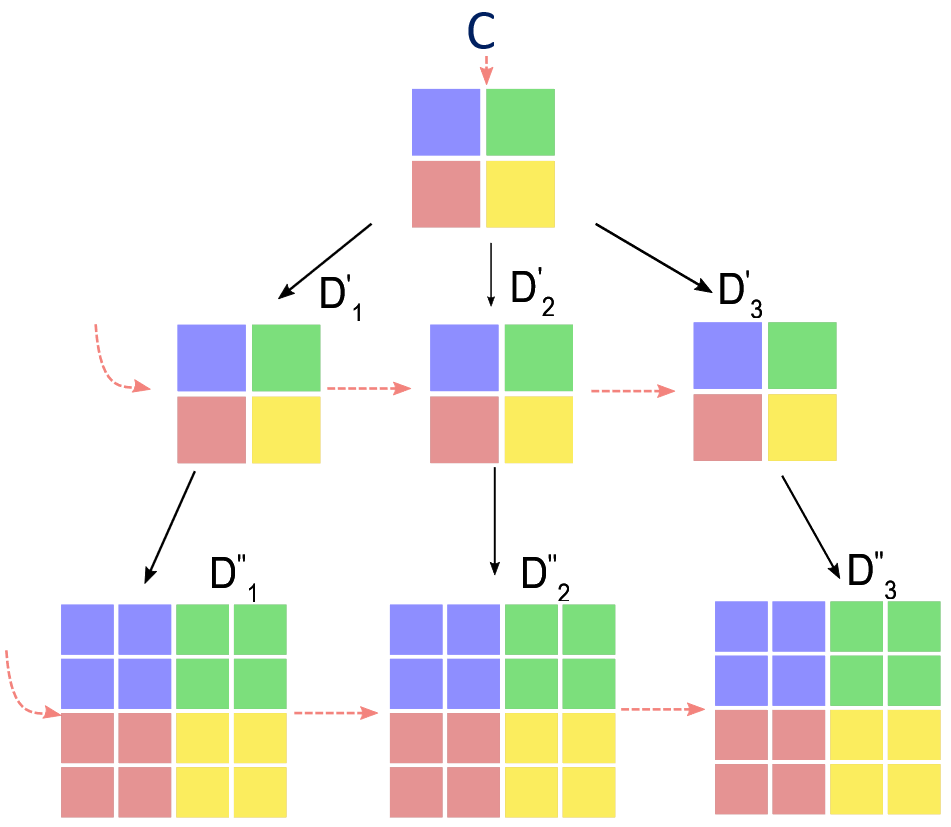}
	\caption[Breadth-first block-coding in BBHex]{Sequence of coding individual sub-bands (or blocks) in BBHex. Here, $C$ is the coarse approximation. For $n = 1, 2, 3$, detail sub-bands $D'_n$ and $D''_n$ correspond to two levels of decomposition via the DHWT.}
	\label{fig:bfs_blockcoding}
	\vspace{-1.0em}
\end{figure}

Our supplementary proposed method is BBHex (breadth-first block-coding for hexagonal images). BBHex is inspired by block-based coding where each sub-band (or block) is encoded separately. Nevertheless, tree-based coding techniques are used in BBHex to encode each sub-band, and the codestream is the concatenation of the significance map of the whole wavelet tree as the sub-bands are passed in breadth-first traversal (see Fig.~\ref{fig:bfs_blockcoding}). Compared to SBHex, the difference is that the individual sub-bands are not mapped into a spiral tree, and the entire tree is not considered for coding. Contrarily, sub-bands of the wavelet tree are individually coded. Coefficients within each sub-band are traversed using our hexagonal scanning order (Sec.~\ref{sec:hexScanning}) and sub-bands are processed in a breadth-first manner. 
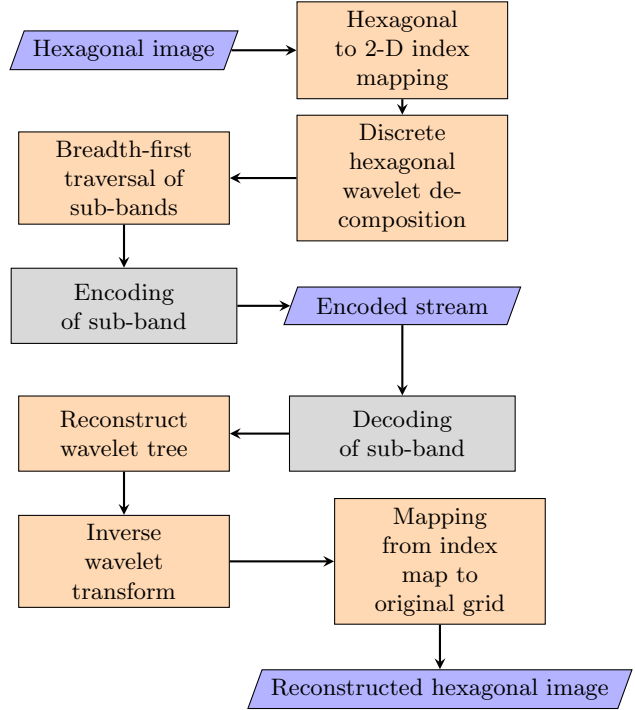
\begin{figure}[h!]
\centering
    \begin{tikzpicture}[node distance=1.7cm]
        \tikzstyle{every node}=[font=\small]
            \node (in1) [io] {Hexagonal image};
            \node (pro1) [process, right of=in1, xshift=2cm] {Hexagonal to 2-D index mapping};
            \node (pro2) [process, below of=pro1] {Discrete hexagonal wavelet decomposition};
            \node (pro3) [process, left of=pro2, xshift=-2cm] {Breadth-first traversal of sub-bands};
            \node (pro4) [process2, below of=pro3] {Encoding of sub-band};
            \node (in2) [io, right of=pro4, xshift=2cm] {Encoded stream};
            \node (pro5) [process2, below of=in2] {Decoding of sub-band};
            \node (pro6) [process, left of=pro5, xshift=-2cm] {Reconstruct wavelet tree};
            \node (pro7) [process, below of=pro6] {Inverse wavelet transform};
            \node (pro8) [process, right of=pro7, xshift=2.5cm] {Mapping from index map to original grid};
            \node (out) [io, below of=pro8] {Reconstructed hexagonal image};

        \draw [arrow] (in1) -- (pro1);
        \draw [arrow] (pro1) -- (pro2);
        \draw [arrow] (pro2) -- (pro3);
        \draw [arrow] (pro3) -- (pro4);
        \draw [arrow] (pro4) -- (in2);
        \draw [arrow] (in2) -- (pro5);
        \draw [arrow] (pro5) -- (pro6);
        \draw [arrow] (pro6) -- (pro7);
        \draw [arrow] (pro7) -- (pro8);
        \draw [arrow] (pro8) -- (out);
\end{tikzpicture}
\caption{Pipeline of BBHex}
\label{fig:pipeline_bbhex}
\vspace{-1.0em}
\end{figure}

The coding-decoding pipeline of BBHex is illustrated in Fig~\ref{fig:pipeline_bbhex}. The encoding also starts with the 2-D index mapping of the given hexagonal image. After the mapping, discrete hexagonal wavelet transformation is performed according to the preferred level of decomposition. To ease the processing, a list of sub-bands is generated by walking in a breadth-first-traversal manner in the tree. Each of the sub-bands is coded using the prescribed parent-to-children relationship as described in Sec.~\ref{sec:parent2childrenreln}. After coding one sub-band, a separator in concatenated in the codestream  and subsequent sub-bands are managed accordingly. 

For the decoding part, the codestream for each sub-band is split using the separator, and a list of codes (1-D array of symbols) is generated. Next, the whole wavelet tree is reconstructed from decoding in a reverse manner, which includes creating the grid of wavelet coefficients for each sub-band and re-organizing them back in the tree. Following the reconstruction of the wavelet tree, inverse wavelet transformation is executed to reconstruct the pixel values in the index map. 

In BBHex, each of the sub-bands is considered individually when testing for insignificant trees. So, the depth of zero trees is less than SBHex; for the magnitude test of the tree, BBHex examines fewer coefficients at each level compared to EZW or SBHex.
On the other hand, BBHex produces smaller zero trees as it is restricted to a specific sub-band levels. Even though it may encode more coefficients into zero-trees, larger trees are not encoded like SBHex; this can lead to slightly more symbols in the codestream depending on the image. The advantage is that the encoding process is faster since fewer coefficients need to be examined when checking for zero-trees.

\section{Results and Discussion}
\label{sec:result_discussions}

In order to investigate the effectiveness of our proposed coding schemes, a varied collection of image sets is tested so as to ensure that the schemes are applied to broad range of images. The categories of test images are as follows:
\begin{itemize}
\item Smooth image: Synthetically generated 2-D chirp~\cite{wiki:chirp}.
\item Non-smooth image: Synthetically generated circular checkerboard texture.
\item General set of true color images: Kodak PhotoCD data set~\cite{kodak:image} which includes natural images, landscape images, face images and photographic images.
\item Astronomical images: NASA Images~\cite{nasa:image}. 
\end{itemize}
In total, 29 (two synthetic and twenty-seven photographic) images are used for testing. 

\subsection{Experimental setup}
For synthetic images, we sample the functions (chirp, checkerboard) on hexagonal and Cartesian grids of equivalent resolutions. 
A common bounding box (holding four corners of the rectangular region of interest) ensures that the same image content is selected for comparison for both the Cartesian and hexagonal grids. Both grids have a coincident origin and the same sampling density in order to ensure that they both contain approximately the same number of pixels.

Natural color images used in the experiments come from test image databases (such as Kodak PhotoCD, NASA images, as mentioned in the datasets). These images are only available in Cartesian format, so an appropriate resampling step is necessary to acquire images on comparable Cartesian and hexagonal grids. The images are downsampled and cropped out according to the region of interest from their high resolution versions. Bicubic interpolation on the original high resolution Cartesian image is used to produce downsampled versions of equivalent quality on the Cartesian and hexagonal grids. We use Keys bicubic interpolation~\cite{keys81} which is a third-order interpolation scheme. This ensures that the downsampling step uses a higher order interpolation scheme compared to the wavelet order used for compression (see below). Since we do not have access to hexagonally sampled benchmark images, this resampling process in unavoidable. It mimics the analog-to-digital conversion process for digital images. The  high  resolution  Cartesian images  when  paired with  Keys cubic  interpolation  serve  as a  surrogate  for  an  analog  signal  from  which appropriately  sampled  Cartesian  and  hexagonal  images  are  obtained. All our  tests  are  conducted  on 362×362 Cartesian grids and 256×256×2 hexagonal grids.

For our hexagonal image compression schemes, DHWT (discrete hexagonal wavelet transform) is performed using the aforementioned second-order filters (Fig.~\ref{fig:dhwtfilters}). On the other hand, Daubechies' second-order wavelet (DB2 with four taps and two vanishing moments) is used to perform the DWT for sub-band coding on the Cartesian grid via EZW. DB2 filters are compactly supported, asymmetric and bi-orthogonal~\cite{daubechies1992ten}. This ensures that both wavelets have two vanishing moments and hence provide a second-order approximation which is lower than the approximation order of the preceding bicubic interpolation step.

In order to compare the quality of reconstruction on the hexagonal grid, both visually and numerically, the reconstructed image coefficients on the hexagonal grid are used to resample the image onto a $362 \times 362$ Cartesian grid. This resampled image is then used for display and comparison purposes. This resampling step is due to the inherent limitation of current display/print technologies that only support Cartesian images. However, it also ensures that all pixel-wise comparisons (see Sec.~\ref{sec:performanceMetrics}) are performed at the same spatial pixel locations. In order to ensure that we keep all errors consistent, triangular (barycentric) linear interpolation is used for this resampling step. The interpolant corresponding to this is the Courant interpolating function which is generated by the low-pass analysis filter (Secion~\ref{sec:multiresolution}).
No resampling is necessary for the images that follow the Cartesian compression and reconstruction pipeline.   

Since the human visual system is much more susceptible to variations in brightness than color, most codecs represent color in the YCbCr space to dedicate more bandwidth towards luminance (Y) rather than chroma (Cb and Cr). For our experiments, the RGB channels of color images are converted into YCbCr space using the transformation proposed by Poynton~\cite{poynton1996technical}. Then only the luminance (Y) space is considered for compression. During computation, if the input image contains uint8, then Y is in the range [16, 235], and Cb and Cr are in the range [16, 240].

  According to empirical evidence, the quality of reconstruction increases with the number of decomposition levels in multiresolution~\cite{pearlman_said_2011}. However, the quality saturates for levels greater than six~\cite{BhokareGanesh2012Ecos}. At smaller numbers, the size of zero-trees is smaller and thus the coding gives less compression gain. As the number of decomposition level increases (level $> 3$), the size of zero-trees starts growing. However, as the levels increase above 6, the number of isolated zeros also increases due to the presence of significant coefficients in the higher frequency bands. Motivated by this, we perform the decomposition to level 6 in all our tests.

\subsection{Performance metrics}
\label{sec:performanceMetrics}
The performance of the proposed sub-band coding schemes is evaluated qualitatively and quantitatively under varying bitrate settings.  In our quantitative evaluation, the reconstruction of data features in the different decomposition levels and progressive transmission are assessed using PSNR and overall-SSIM. SSIM (Structural Similarity Index Metric)~\cite{ZhouWang2004Iqaf} computes the local patterns of normalized pixel intensities for luminance and contrast. This metric has been specifically designed to match the human visual system. In current literature, it has been established to be significantly more representative than losses in the $l_p$ family and variants of PSNR~\cite{waveone2017}. However, since PSNR is also a widely used metric, we report both SSIM and PSNR results. All SSIM and PSNR computations are performed on a $362 \times 362$ Cartesian grid as explained in the previous section. 

For our experiments, BPP (bits-per-pixel) indicates the average number of bits used to represent the Y-channel. This space-saving criterion is defined as the ratio of the size (in bits) of the reduced data to the number of \emph{coefficients} in the original grid. For example, for the $256 \times 256 \times 2$ hexagonal input, BPP = number of bits in codestream / ($256 \times 256 \times 2$), whereas for the $362 \times 362$ Cartesian images, BPP = number of bits in codestream / ($362 \times 362$).

\begin{figure*}[t]
  \centering
  \begin{subfigure}[c]{0.22\textwidth}\raggedleft
    \raisebox{-.5\height}{\includegraphics[width=.9\textwidth]
    {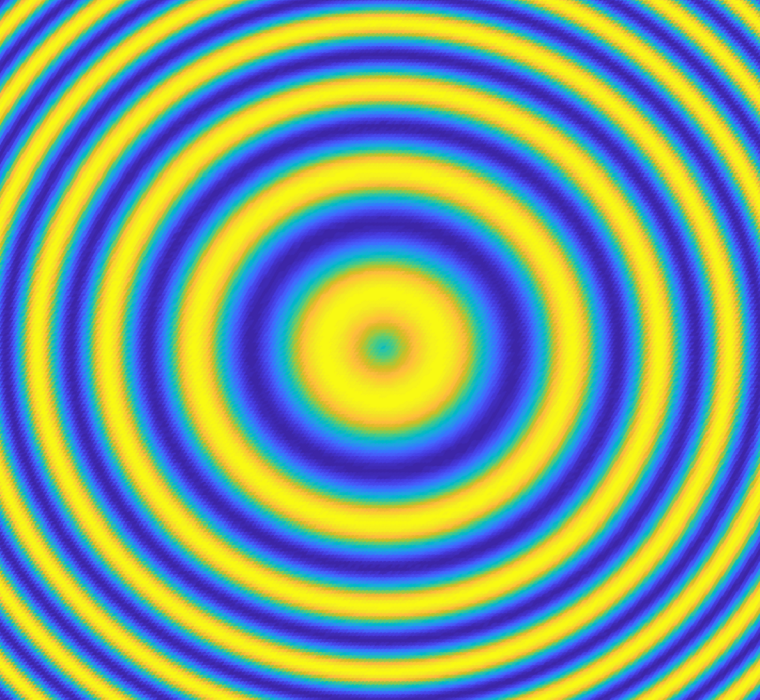}}
    \caption*{Chirp}
    \raisebox{-.5\height}{\includegraphics[width=.9\textwidth]
    {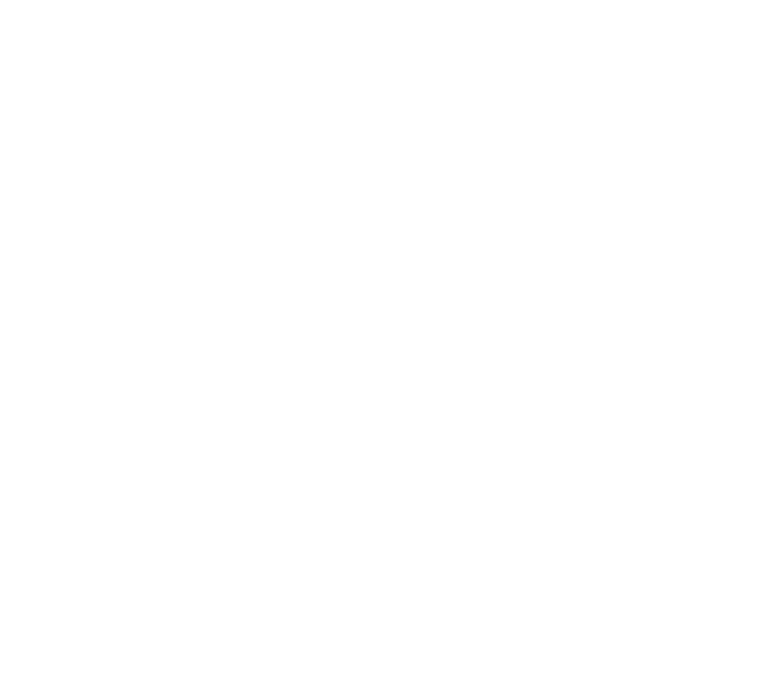}}
    \caption*{        }
    \raisebox{-.5\height}{\includegraphics[width=.9\textwidth]
    {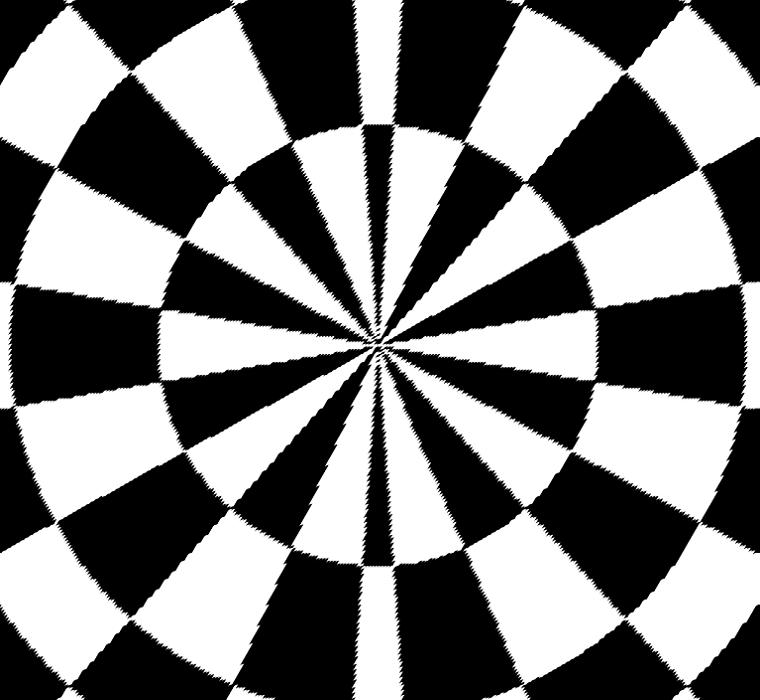}}
    \caption*{Checkerboard}
    \raisebox{-.5\height}{\includegraphics[width=.9\textwidth]
    {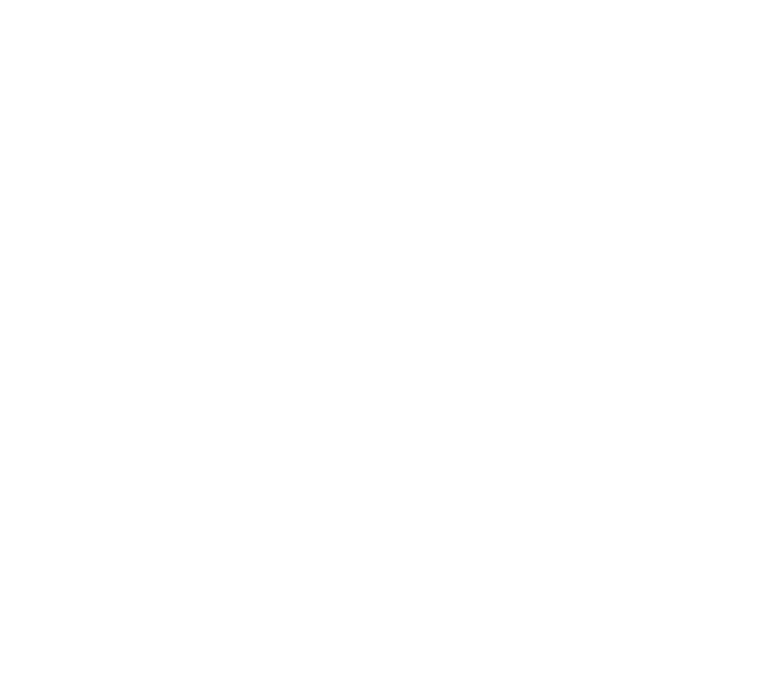}}
    \caption*{       }
    \caption{Original}
\end{subfigure}%
\hspace{0.3em}
\begin{subfigure}[c]{0.22\textwidth}\raggedleft
    \includegraphics[width=.9\textwidth]  
    {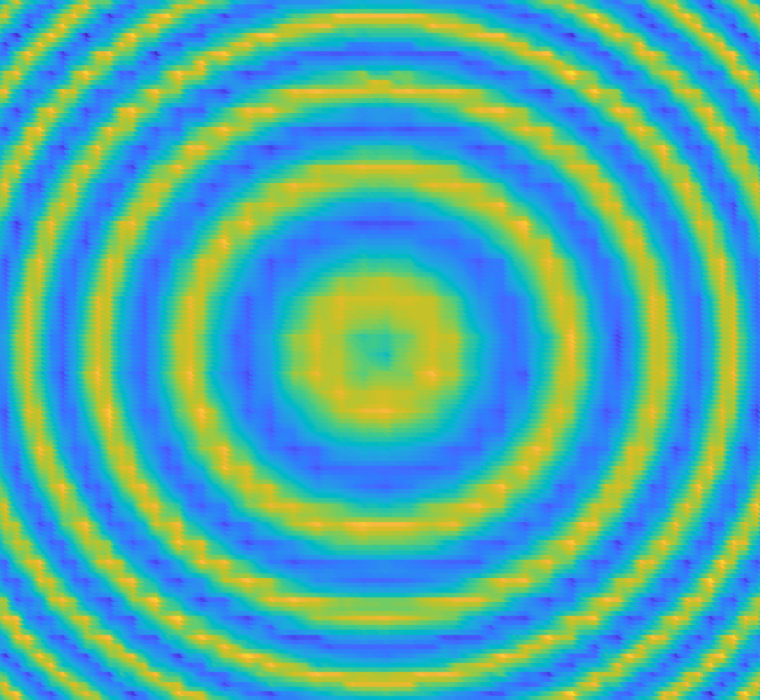}
    \caption*{\fontsize{7}{8}\selectfont BPP:0.5, PSNR:20.08 
    }
    \includegraphics[width=.9\textwidth]
    {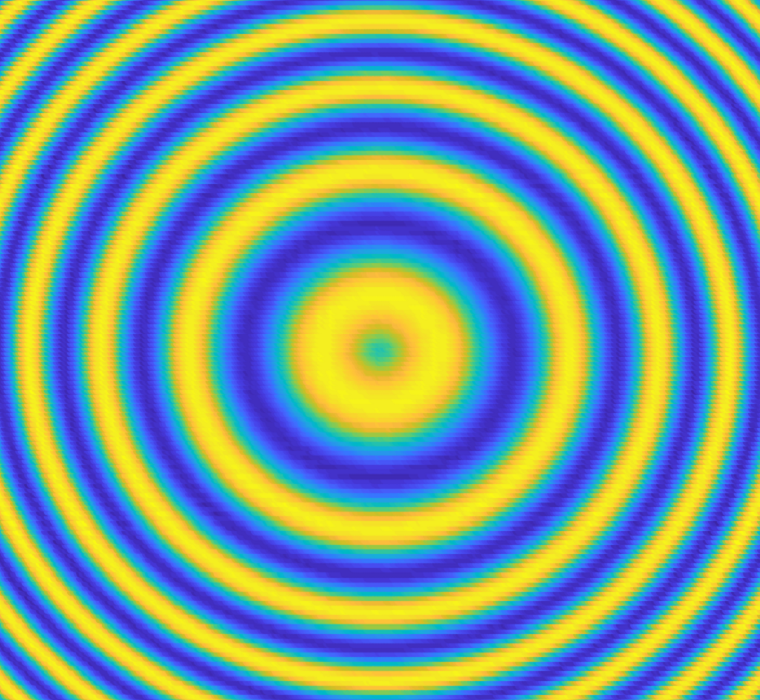}
    \caption*{\fontsize{7}{8}\selectfont BPP:2.5,  PSNR:92.34 
    }
    \includegraphics[width=.9\textwidth]
    {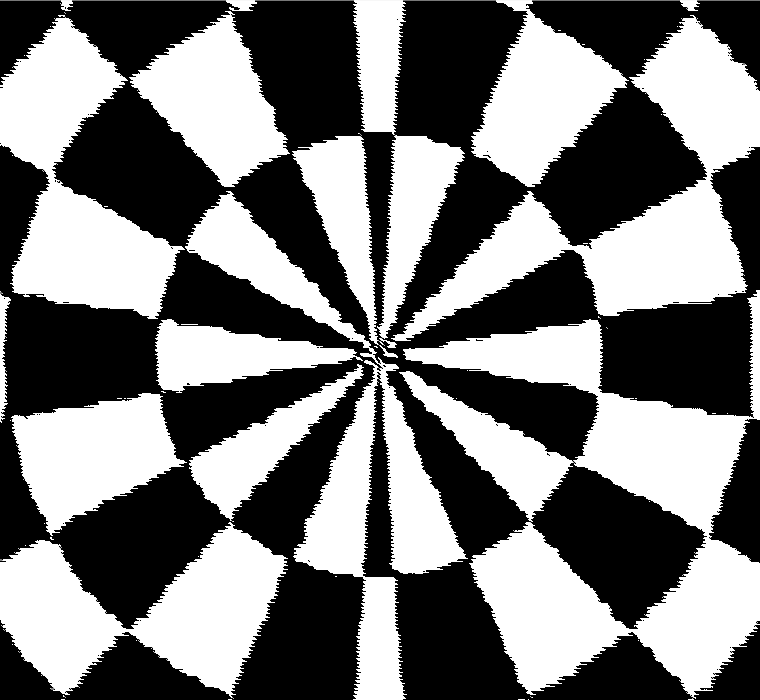}
    \caption*{\fontsize{7}{8}\selectfont BPP:0.2, PSNR:32.87 
    }
    \includegraphics[width=.9\textwidth]
    {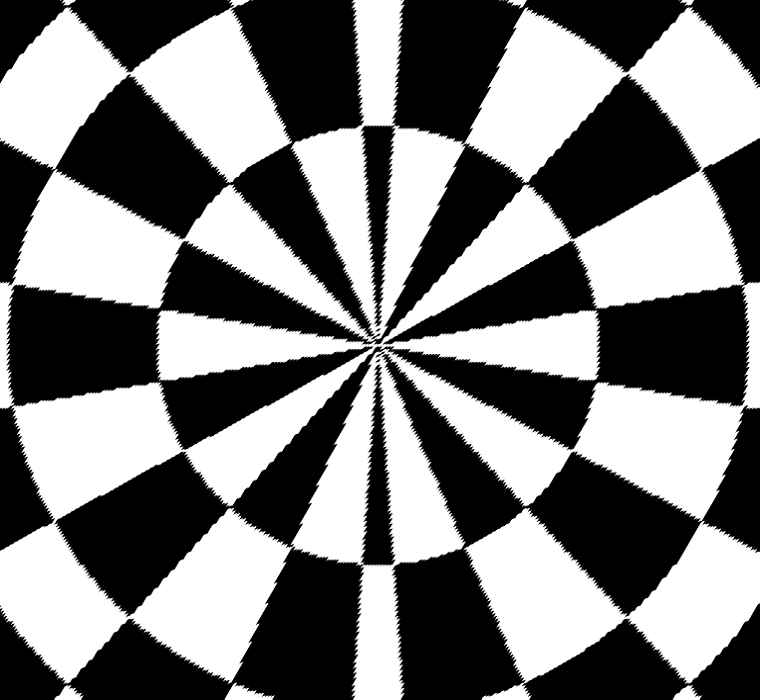}
    \caption*{
    \fontsize{7}{8}\selectfont 
    BPP:3.8, PSNR:91.57 
    }
    \caption{EZW (DB2)}
\end{subfigure}
\hspace{0.3em}
\begin{subfigure}[c]{0.22\textwidth}\centering
    \includegraphics[width=.9\textwidth]  
    {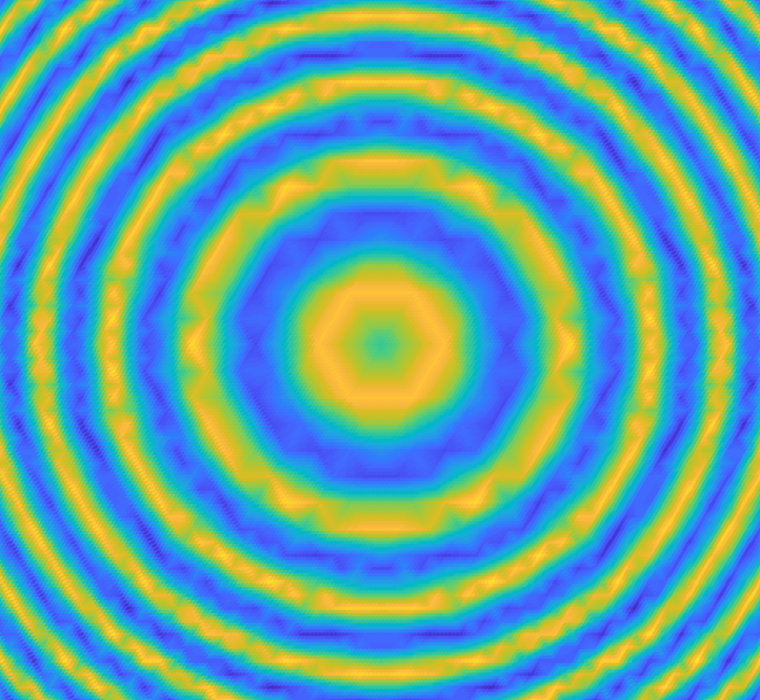}
    \caption*{\fontsize{7}{8}\selectfont BPP:.12, PSNR:47.54 
    }
    \includegraphics[width=.9\textwidth]
    {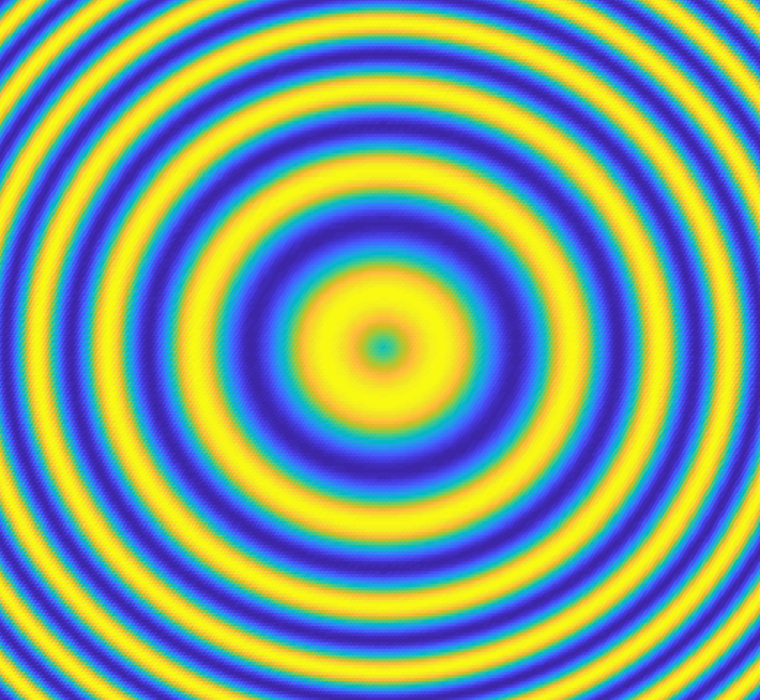}
    \caption*{\fontsize{7}{8}\selectfont BPP:1.0, PSNR:95.31 
    }
    \includegraphics[width=.9\textwidth]
    {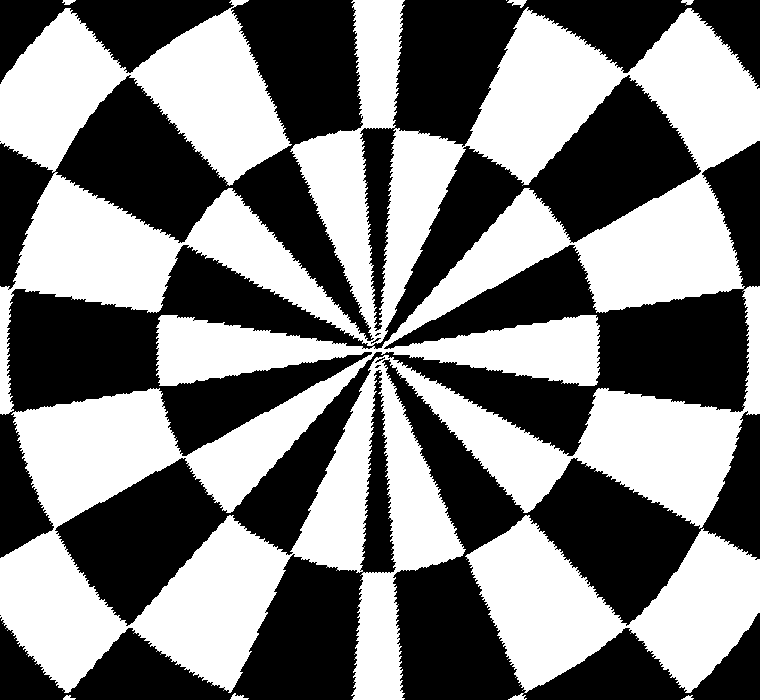}
    \caption*{\fontsize{7}{8}\selectfont BPP:0.1,   PSNR:52.23 
    }
    \includegraphics[width=.9\textwidth]
    {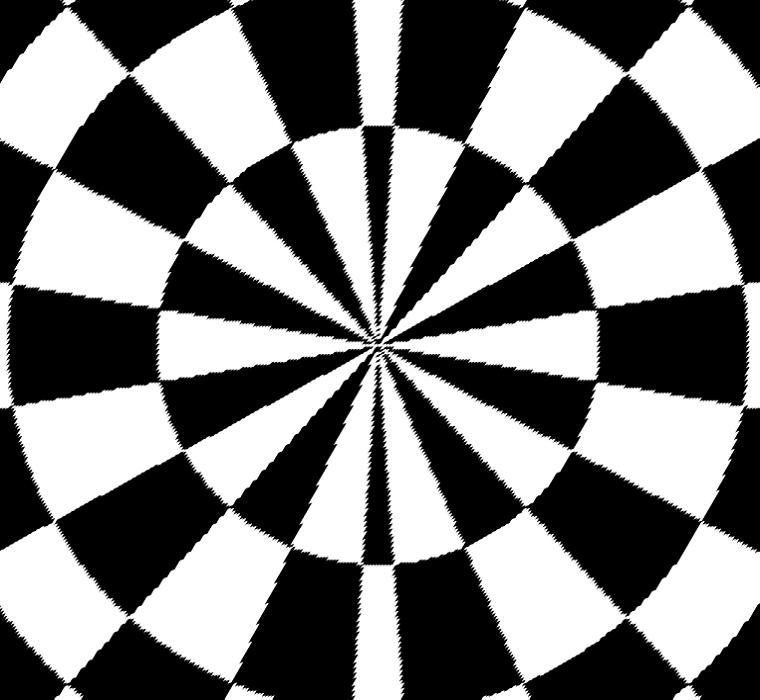}
    \caption*{\fontsize{7}{8}\selectfont BPP:1.9, PSNR:98.12 
    }
    \caption{SBHex}
\end{subfigure}
\hspace{0.3em}
\begin{subfigure}[c]{0.22\textwidth}\centering
     \includegraphics[width=.9\textwidth]  
    {results/chirp/dhwt_hex_6/7}
    \caption*{\fontsize{7}{8}\selectfont BPP:.14, PSNR:47.54 
    }
    \includegraphics[width=.9\textwidth]
    {results/chirp/dhwt_hex_6/10}
    \caption*{\fontsize{7}{8}\selectfont BPP:1.2, PSNR:95.31 
    }
     \includegraphics[width=.9\textwidth]
     {results/checkerBoard/dhwt_hex_6/6}
    \caption*{\fontsize{7}{8}\selectfont BPP:.74, PSNR:52.23 
    }
    \includegraphics[width=.9\textwidth]
    {results/checkerBoard/dhwt_hex_6/14}
    \caption*{\fontsize{7}{8}\selectfont BPP:1.8, PSNR:98.12 
    }
    \caption{BBHex}
\end{subfigure}

\caption[Qualitative comparison of synthetic images]{Qualitative comparison of coding schemes (b) EZW (c) SBHex and (d) BBHex for synthetic images. ``Chirp'' (colormapped with \textit{parula})and ``Checkerboard'' are compressed to different BPP values for hexagonal lattice (size: $256 \times 256 \times 2$) and equivalent Cartesian grid (size: $362 \times 362$). For hexagonal and Cartesian grid, DHWT filters and DB2 are used respectively. The multiresolution decomposition level is 6.}
\label{fig:synthetic_image_result}
\vspace{-1.0em}
\end{figure*}
 
\subsection{Qualitative Results}
\begin{figure*}[t]
  \centering
  \begin{subfigure}[c]{0.22\textwidth}\raggedleft
   
    \raisebox{-.3\height}{\includegraphics[width=.9\textwidth]
    {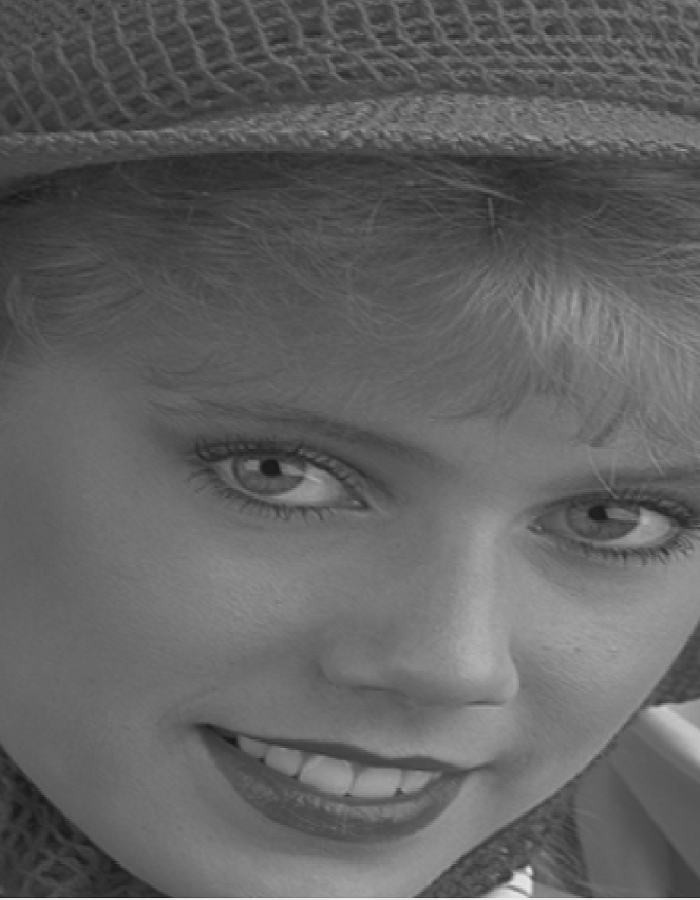}}
    \caption*{Red Hat}
    
    \raisebox{-.3\height}{\includegraphics[width=.9\textwidth]
    {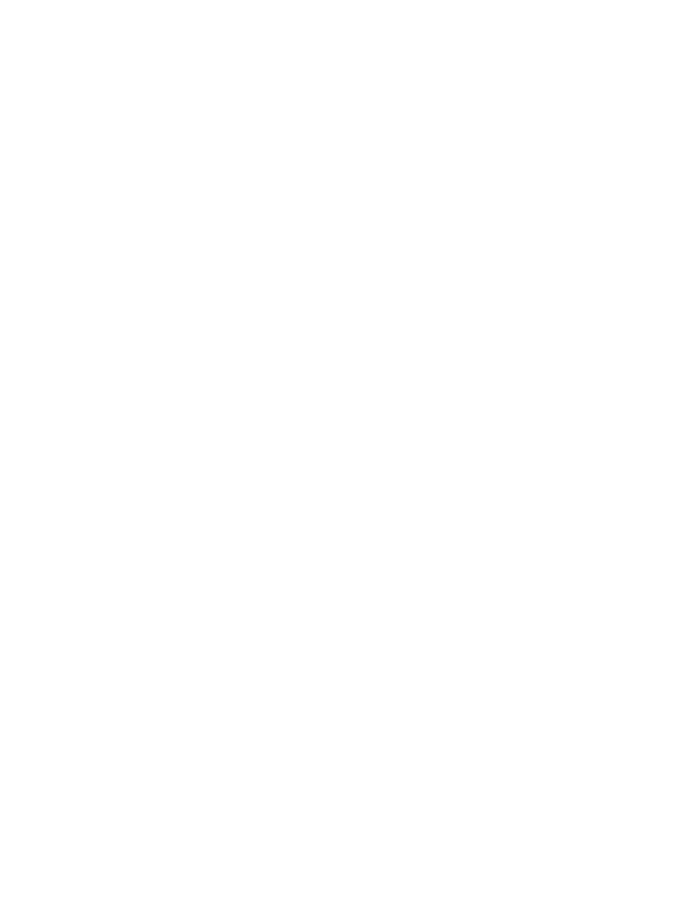}}
    \caption*{}
    \caption{Original}
\end{subfigure}%
\hspace{0.1em}
\begin{subfigure}[c]{0.22\textwidth}\raggedleft
    
    \includegraphics[width=.9\textwidth]
     {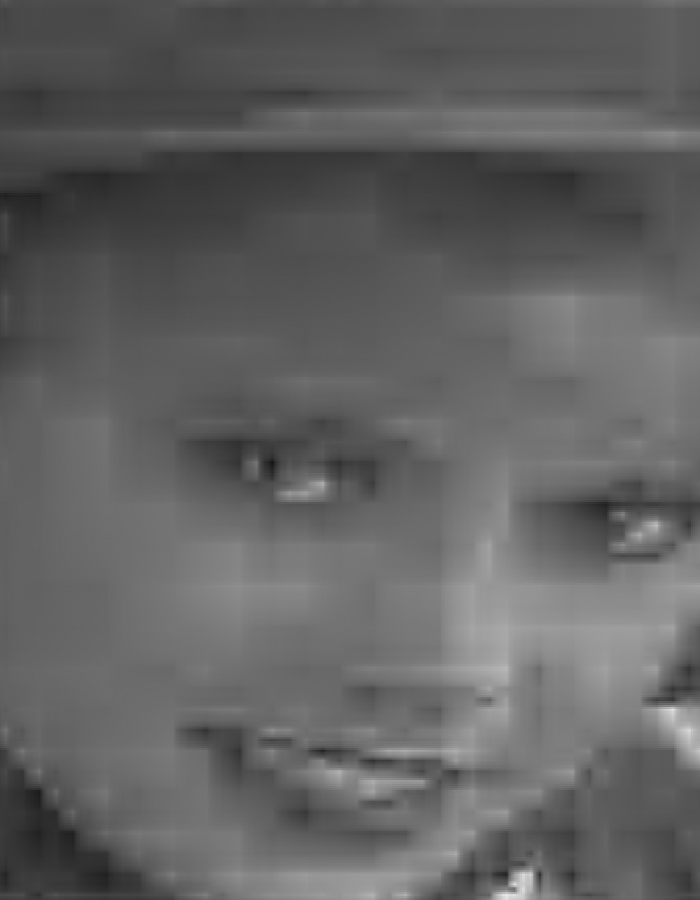}
    \caption*{\fontsize{7}{8}\selectfont BPP:0.2, PSNR:21.14 
    }
    \includegraphics[width=.9\textwidth]
    {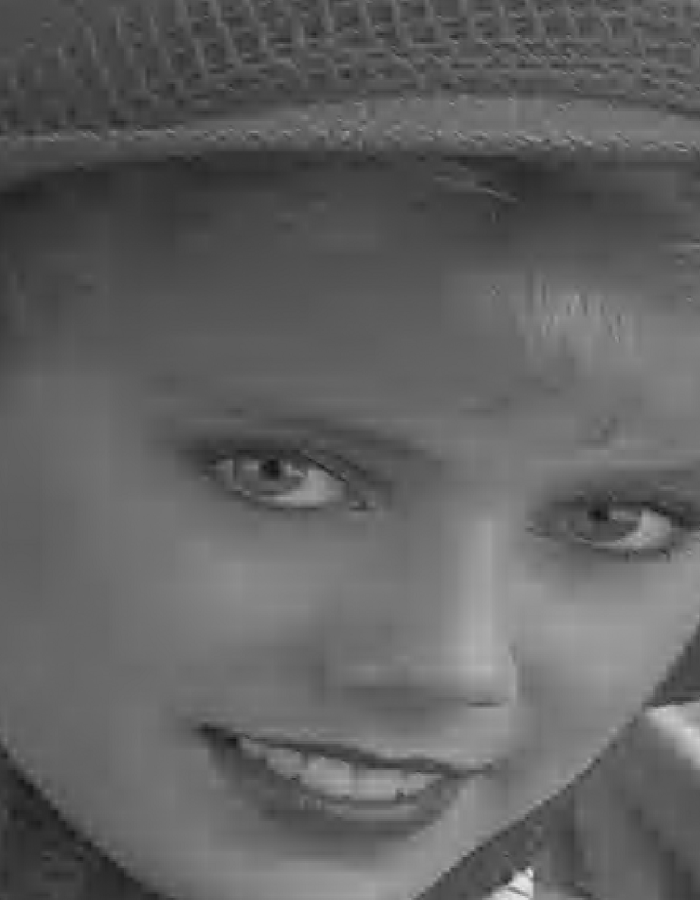}
    \caption*{\fontsize{7}{8}\selectfont BPP:1.8, PSNR:38.78 
    }
    \caption{EZW (DB2)}
\end{subfigure}
\hspace{0.5em}
\begin{subfigure}[c]{0.22\textwidth}\centering
  
    \includegraphics[width=.9\textwidth]
    {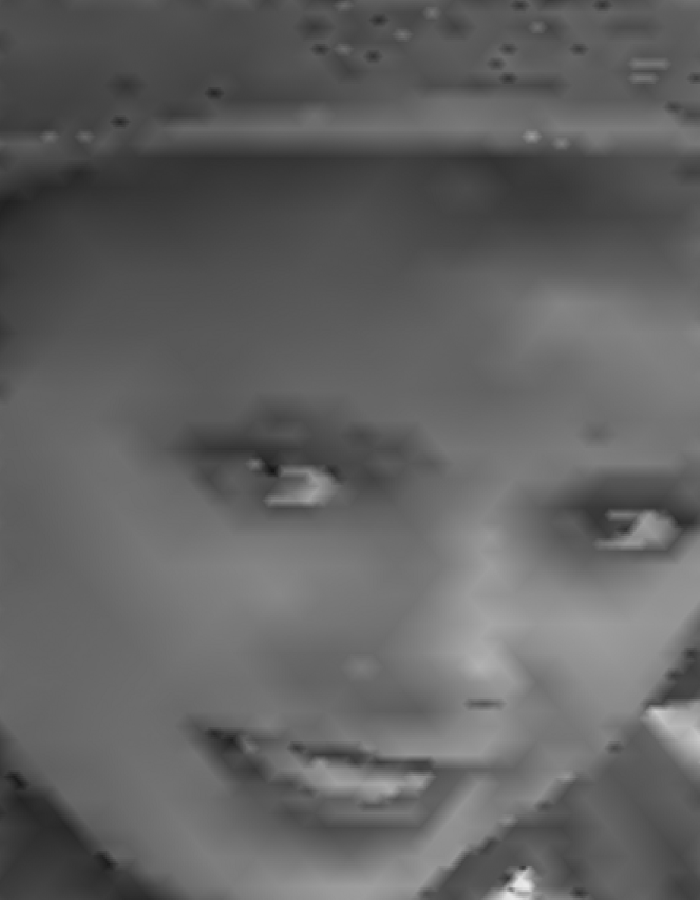}
    \caption*{\fontsize{7}{8}\selectfont BPP:0.14, PSNR:33.7 
    }
    \includegraphics[width=.9\textwidth]
    {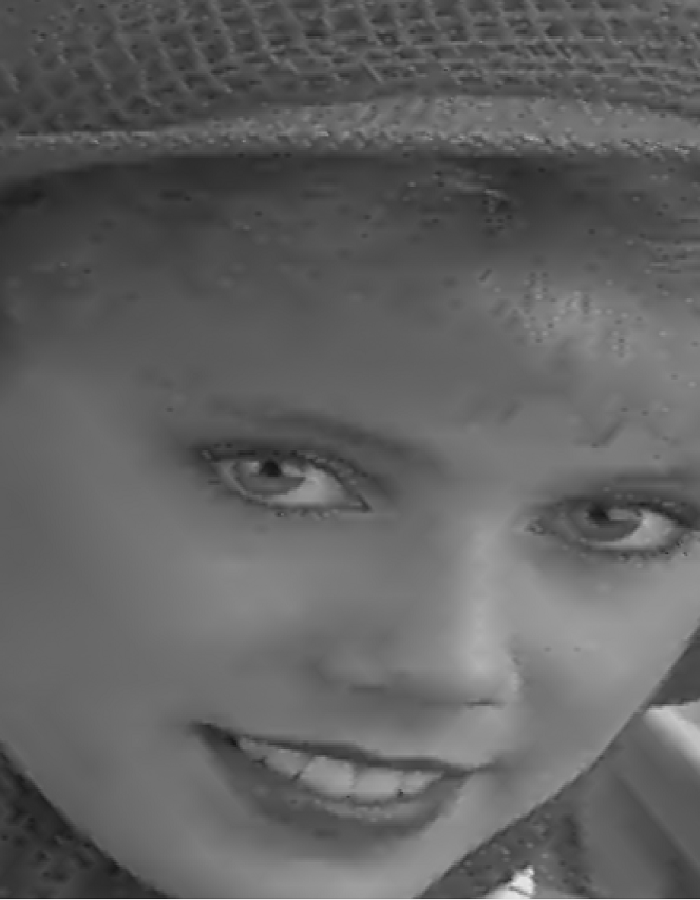}
    \caption*{\fontsize{7}{8}\selectfont BPP:0.6, PSNR:58.21 
    }
    \caption{SBHex}
\end{subfigure}
\hspace{0.5em}
\begin{subfigure}[c]{0.22\textwidth}\centering
   
    \includegraphics[width=.9\textwidth]
    {results/kodim04/dhwt_hex_6/8}
    \caption*{\fontsize{7}{8}\selectfont BPP:0.18, PSNR:33.7 
    }
    \includegraphics[width=.9\textwidth]
   {results/kodim04/dhwt_hex_6/10}
    \caption*{\fontsize{7}{8}\selectfont BPP:0.7, PSNR:58.21 
    }
    \caption{BBHex}
\end{subfigure}

\caption[Qualitative comparison of face image]{Qualitative comparison of coding schemes (b) EZW (c) SBHex and (d) BBHex for face image ``Red hood'' from Kodak PhotoCD dataset. The image is compressed to different BPP values for hexagonal lattice (size: $256 \times 256 \times 2$) and equivalent Cartesian grid (size: $362 \times 362$). For hexagonal and Cartesian grid, DHWT filters and DB2 are used respectively. The multiresolution decomposition level is 6.
}
\label{fig:kodak_result1}
\vspace{-1.0em}
\end{figure*}

\begin{figure*}[t]
  \centering
  \begin{subfigure}[c]{0.22\textwidth}\raggedleft
   
    \raisebox{-.3\height}{\includegraphics[width=.9\textwidth]
    {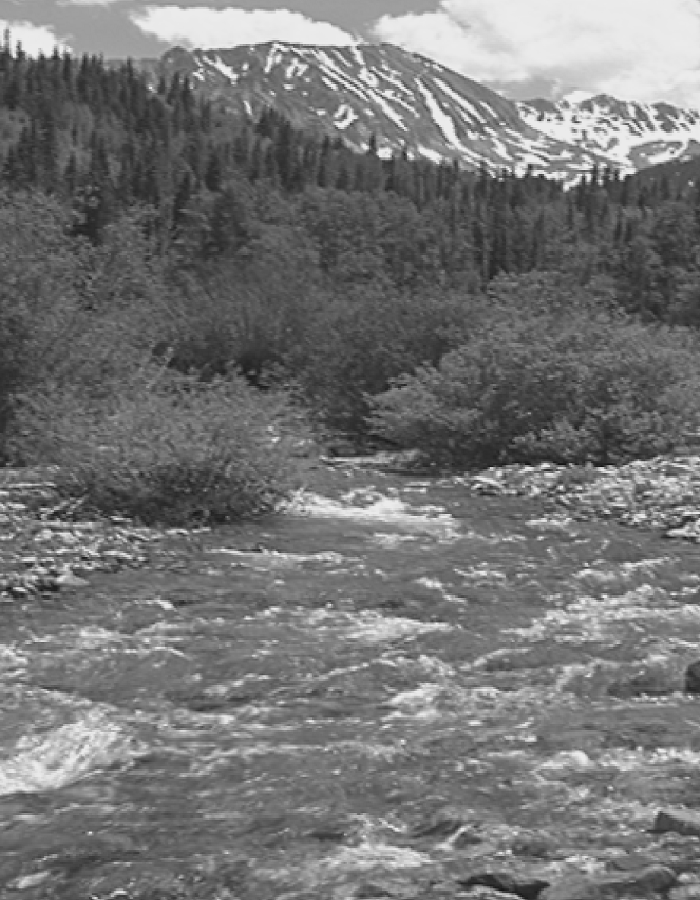}}
    \caption*{Mountain stream}
    
    \raisebox{-.3\height}{\includegraphics[width=.9\textwidth]
    {results/blank_image_700x900}}
    \caption*{}
    \caption{Original}
\end{subfigure}%
\hspace{0.1em}
\begin{subfigure}[c]{0.22\textwidth}\raggedleft
    
    \includegraphics[width=.9\textwidth]
     {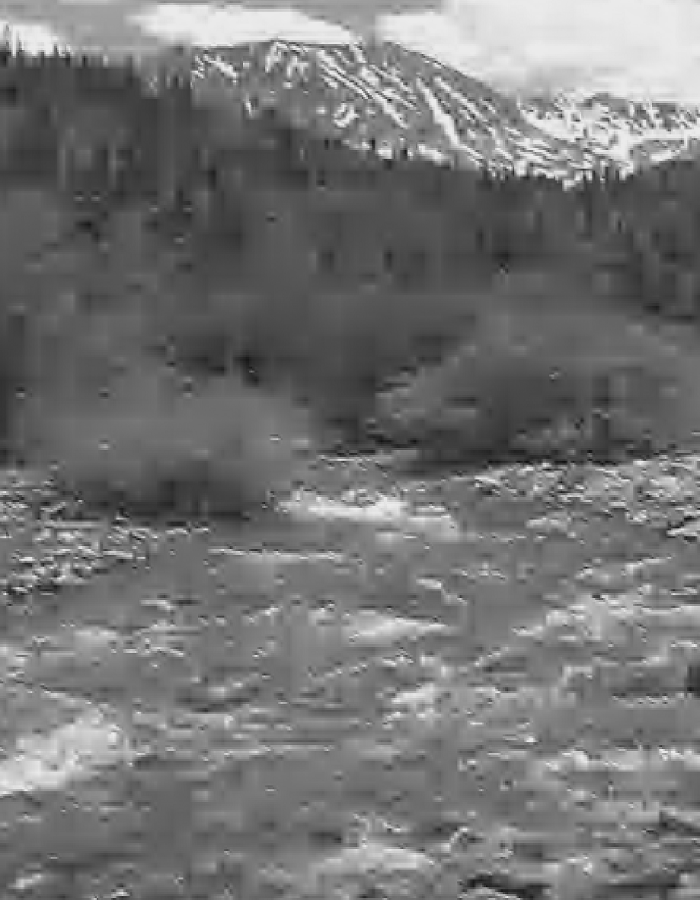}
    \caption*{\fontsize{7}{8}\selectfont BPP:0.9, PSNR:21.14
    }
    \includegraphics[width=.9\textwidth]
    {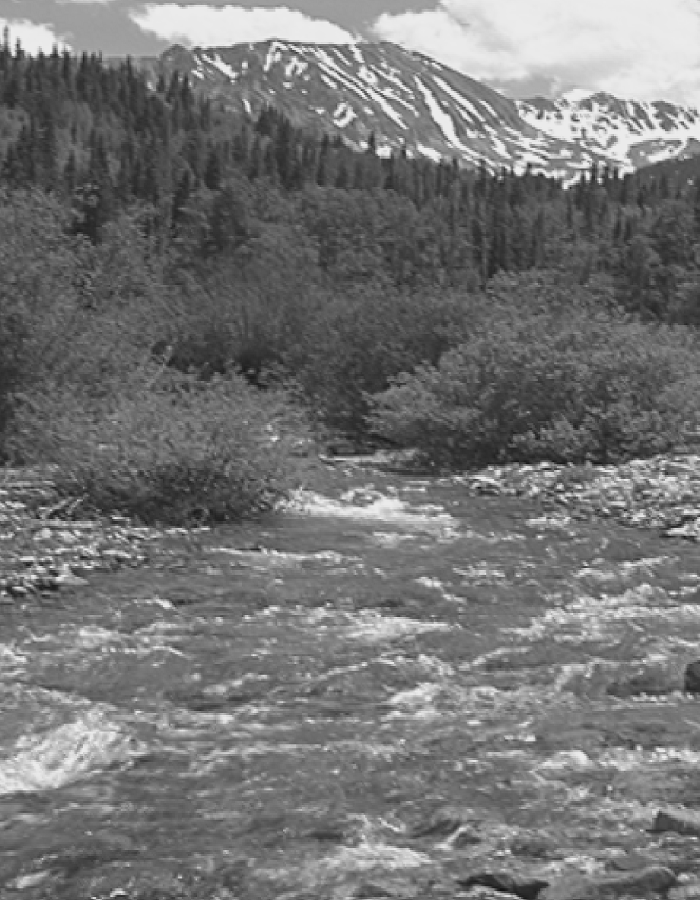}
    \caption*{\fontsize{7}{8}\selectfont BPP:0.5, PSNR:34.7
    }
    \caption{EZW (DB2)}
\end{subfigure}
\hspace{0.5em}
\begin{subfigure}[c]{0.22\textwidth}\centering
  
    \includegraphics[width=.9\textwidth]
    {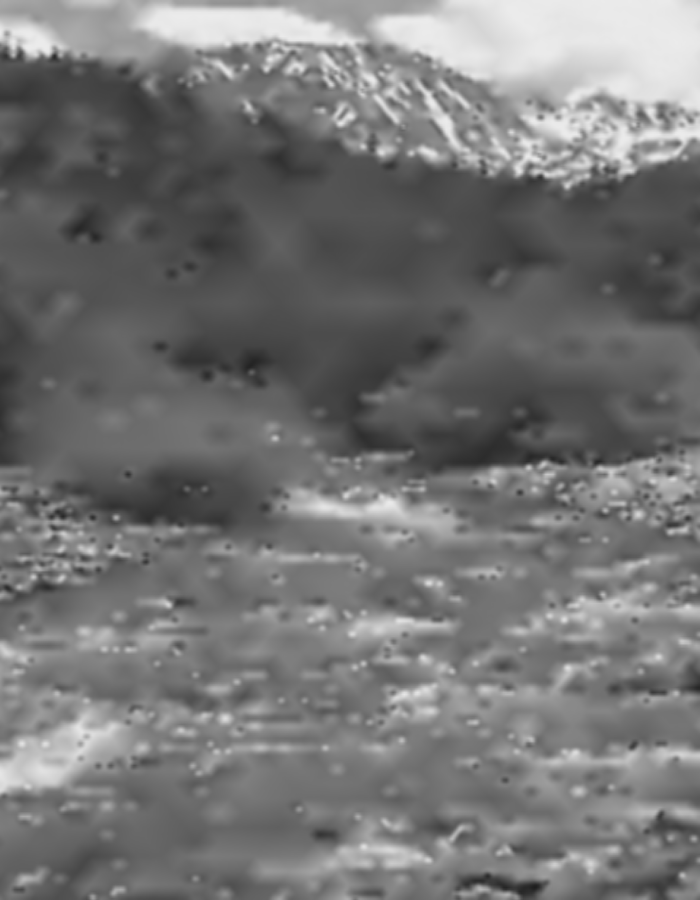}
    \caption*{\fontsize{7}{8}\selectfont BPP:0.2, PSNR:34.5 
    }
    \includegraphics[width=.9\textwidth]
    {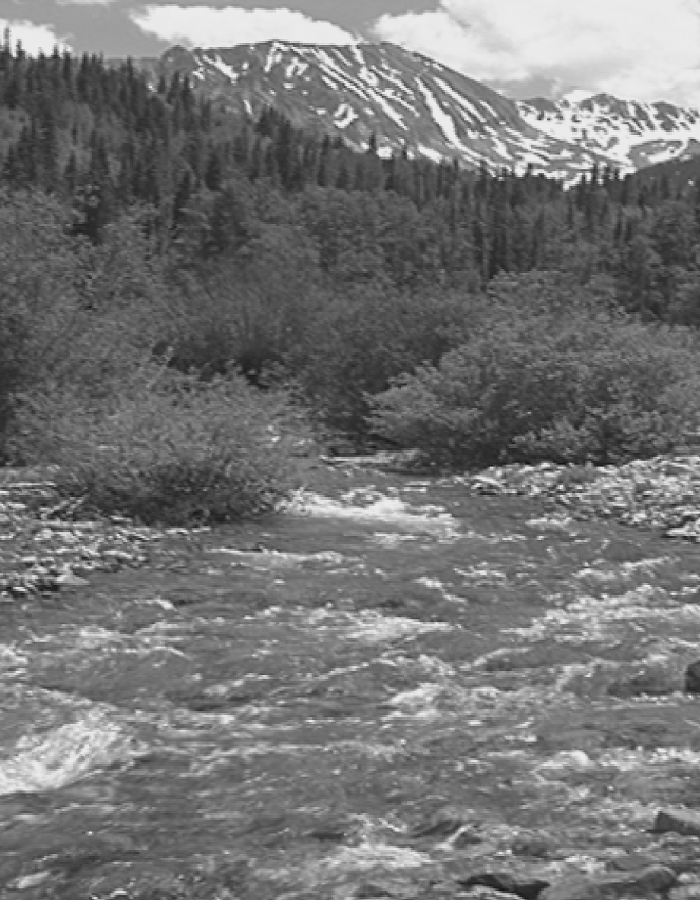}
    \caption*{\fontsize{7}{8}\selectfont BPP:0.47, PSNR:43.2 
    }
    \caption{SBHex}
\end{subfigure}
\hspace{0.5em}
\begin{subfigure}[c]{0.22\textwidth}\centering
   
    \includegraphics[width=.9\textwidth]
    {results/kodim13/dhwt_hex_6/9}
    \caption*{\fontsize{7}{8}\selectfont BPP:0.18, PSNR:34.5
    }
    \includegraphics[width=.9\textwidth]
   {results/kodim13/dhwt_hex_6/12}
    \caption*{\fontsize{7}{8}\selectfont BPP:0.48, PSNR:43.2
    }
    \caption{BBHex}
\end{subfigure}

\caption[Qualitative comparison of natural image]{Qualitative comparison of coding schemes (b) EZW (c) SBHex and (d) BBHex for natural image ``Mountain stream'' from Kodak PhotoCD dataset. The image is compressed to different BPP values for hexagonal lattice (size: $256 \times 256 \times 2$) and equivalent Cartesian grid (size: $362 \times 362$). For hexagonal and Cartesian grids, DHWT filters and DB2 are used, respectively. The multiresolution decomposition level is 6.
}
\label{fig:kodim13_result}
\vspace{-1.0em}
\end{figure*}

The quality of the reconstructed image is directly proportional to the number of bits in the codestream irrespective of the grid used. From the visual examples of reconstruction in Fig.~\ref{fig:synthetic_image_result} and Fig.~\ref{fig:kodak_result1} and ~\ref{fig:kodim13_result}, it is observed that reconstructed images using a reduced symbol stream have severe loss of details and exhibit blocking artifacts. On the other hand, in the reconstructed images with more BPP, the compression artifacts are difficult to notice.

From the visual comparison, we can see the effect of DWT and DHWT on the reconstructed images. At lower BPP values, the reconstruction using DWT compared to DHWT exhibits different compression artifacts. EZW (backed by DB2) on the Cartesian grid produces more of a discontinuous blocking effect. On the other hand, SBHex and BBHex (backed by DHWT) on the hexagonal grid produce clearer images compared to EZW even at low BPP values. For the (synthetically generated) isotropic images (see Fig.~\ref{fig:synthetic_image_result}), SBHex and BBHex on the hexagonal grid also produce better quality as compared to EZW. This is further corroborated by the PSNR values; at lower BPP values, our algorithms yields higher PSNR values.  For the `Red Hat' image (Fig.~\ref{fig:kodak_result1}), the reconstruction of the face at low BPP values is visually more pleasing for SBHex and BBHex compared to EZW. The non-smooth texture details in the hat are also visible in SBHex and BBHex. 

\subsection{Quantitative Results}
PSNR values for various natural images at different bitrates are tabulated in Table~\ref{table:psnr-bpp-table}. For each image, SBHex, BBHex and EZW are compressed to achieve approximately the same PSNR value. For comparison, JPEG2000 is compressed to the same target bit-rate (approximately) as that achieved by SBHex and BBHex, and the corresponding PSNR value is determined. The table shows that both SBHex and BBHex produce superior results than EZW as they can produce similar image quality (PSNR) at much lower bit rates (about 40\%-50\% less). For the same target bit-rate, JPEG2000 produces the best results but it is interesting to note that SBHex and BBHex are very close in comparison. The superiority of JPEG2000 is expected as it uses higher-order wavelets and incorporates tiling. We envisage that by incorporating these strategies in SBHex and BBHex in future, we can further improve the performance of our hexagonal compression schemes.

\begin{table}[h!]
\centering
\caption{Comparison of embedding capacity (bpp) and image quality (PSNR)}
\label{table:psnr-bpp-table}
\begin{small}
\begin{tabular}{ |p{1.5cm}||p{0.9cm}|p{0.9cm}|p{0.9cm}|p{0.9cm}|p{0.9cm}|p{0.9cm}|p{0.9cm}|p{0.9cm}|  }
 \hline
 \hline
 {Images} & \multicolumn{2}{|c|}{JPEG2000} & \multicolumn{2}{|c|}{EZW (DB2)} & \multicolumn{2}{|c|}{SBHex} & \multicolumn{2}{|c|}{BBHex}\\

& bit rate (bpp) & PSNR (dB) & bit rate (bpp) & PSNR (dB) & bit rate (bpp) & PSNR (dB) & bit rate (bpp) & PSNR (dB)\\
 \hline
Lena & 0.0124 & 30.6856 & 0.0226 & 28.4833 & 0.0123 & 28.6081 & 0.0124 & 28.6081\\
 \hline
 Barbara & 0.0117 & 23.7667 & 0.0238 & 22.4671 & 0.0117 & 22.9856 & 0.0118 & 22.9856\\
  \hline
Blackhole & 0.0122 & 52.3667 & 0.0256 & 51.3533 & 0.0120 & 52.6185 & 0.0121 & 52.6185\\
 \hline
 Barnard68 & 0.0122 & 24.0894 & 0.0223 & 23.9732 & 0.0121 & 23.1850 & 0.0122 & 23.1850\\
\hline
 Horsehead & 0.0154 & 30.4047 & 0.0323 & 27.9739 & 0.0154 & 28.9534 & 0.0155 & 28.9534\\
\hline
 NGC602 & 0.0134 & 21.2692 & 0.0222 & 19.9273 & 0.0134 & 20.2315 & 0.0135 & 20.2315\\
\hline
Peppers & 0.0120 & 37.6078 & 0.3224 & 35.8073 & 0.1792 & 36.1852 & 0.1794 & 36.1852\\
\hline
Building & 0.0126 & 25.4610 & 0.0232 & 23.2376 & 0.0124 & 24.8780& 0.0125 & 24.8780\\
\hline
Door & 0.0121 & 32.2971 & 0.0223 & 30.2475 & 0.0120 & 31.7076 & 0.0121 & 31.7076\\
\hline
Hats & 0.0119 & 34.1086 & 0.0218 & 32.8954 & 0.0119 & 33.1448 & 0.0120 & 33.1448\\
\hline
Bikes & 0.0189 & 34.1224 & 0.0221 & 32.8748 & 0.0180 & 33.4922 & 0.0190 & 33.4922\\
\hline
Boat & 0.0120 & 27.9797 & 0.0223 & 25.9054 & 0.0120 & 26.3407 & 0.0121 & 26.3407\\
\hline
Windows & 0.0155 & 31.5824 & 0.0225 & 29.7953 & 0.0151 & 30.6210 & 0.0153 & 30.6210\\
\hline
Market & 0.0110 & 23.5780 & 0.0223 & 21.9567 & 0.0113 & 22.8215 & 0.0114 & 22.8215\\
\hline
Sailboats & 0.0131 & 33.7574 & 0.0220 & 31.6578 & 0.0132 & 32.7290 & 0.0133 & 32.7290\\
\hline
Pier & 0.0121 & 29.0536 & 0.0321 & 27.6947 & 0.0121 & 27.1850 & 0.0122 & 27.1850\\
\hline
Couples & 0.0158 & 33.9963 & 0.0224 & 31.7483 & 0.0158 & 32.0467 & 0.0159 & 32.0467\\
\hline
Rafters & 0.0167 & 27.8875 & 0.0289 & 25.3756 & 0.0169 & 26.1750 & 0.0170 & 26.1750\\
\hline
Tropical & 0.0138 & 37.1417 & 0.0273 & 35.2098 & 0.0139 & 36.1085 & 0.1408 & 36.1085\\
\hline
Stephen & 0.0110 & 32.0215 & 0.0243 & 29.8273 & 0.0110 & 30.9296 & 0.0111 & 30.9296\\
\hline
Model & 0.0148 & 26.6409 & 0.0283 & 24.6675 & 0.0147 & 25.0175 & 0.0148 & 25.0175\\
\hline
Lighthouse & 0.0134 & 30.1367 & 0.0241 & 27.5978 & 0.0133 & 28.9645 & 0.0134 & 28.9645\\
\hline
Plane & 0.0121 & 32.6734 & 0.0273 & 30.9643 & 0.0120 & 31.9589 & 0.0121 & 31.9589\\
\hline
Headlight & 0.0189 & 30.5329 & 0.0321 & 26.9678 & 0.0187 & 27.1232 & 0.0188 & 27.1232\\
\hline
Barn & 0.0129 & 29.3777 & 0.0223 & 27.7467 & 0.0127 & 28.7658 & 0.0128 & 28.7658\\
\hline
Macaws & 0.0120 & 35.7584 & 0.0256 & 33.6785 & 0.0120 & 34.8308 & 0.0121 & 34.8308\\
\hline
Chalet & 0.0178 & 27.2529 & 0.0220 & 24.5378 & 0.0177 & 25.2964 & 0.0179 & 25.2964\\
\hline
\end{tabular}
\end{small}
\end{table}

\begin{figure*}[!ht]
\begin{subfigure}[b]{.5\textwidth}
    \centering
    \includegraphics[scale=0.6]{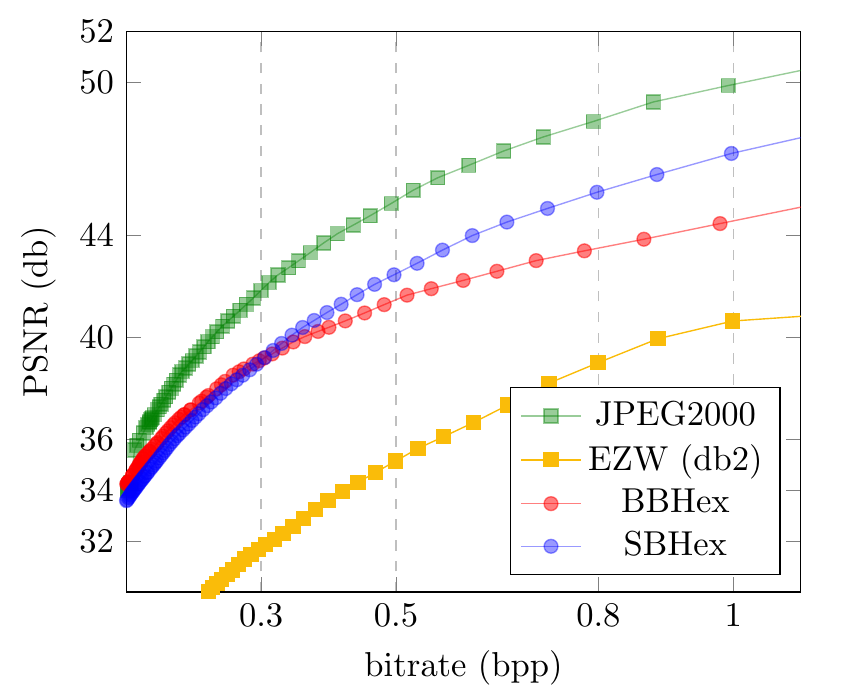}
    \caption{Rate distortion for \textit{Mountain stream} image}
    \label{fig:bpp-psnr-curve-mountain-stream}
\end{subfigure}
 \begin{subfigure}[b]{.5\textwidth}
  \centering
    \includegraphics[scale=.6]{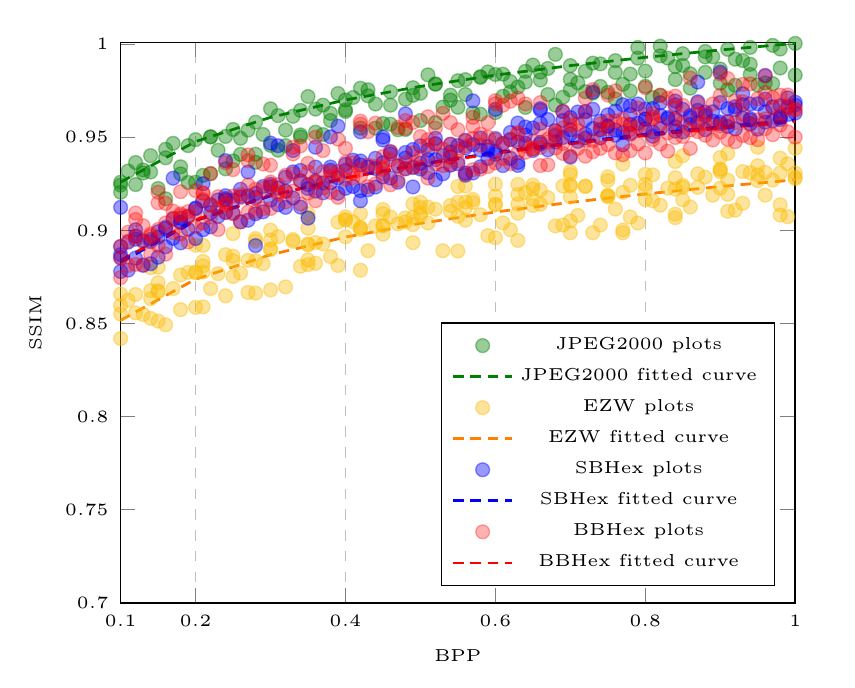}
    \caption{Regression curves}
    \label{fig:fitting_curve_ezw}
 \end{subfigure}
  \hfill
    \caption{Comparative evaluation (rate-distortion curves) for EZW, SBHex and BBHex. (a) PSNR vs. BPP curves for Barbara image compressed to different quality levels. (b) Hill regression curves for all the test images.}
    \label{fig:ezw_results}
    \vspace{-1.0em}
\end{figure*}

To further illustrate the effect of quality deterioration with decreasing bit rate, Fig.~\ref{fig:bpp-psnr-curve-mountain-stream} compares SBHex, BBHex, and EZW with JPEG2000 for the \textit{Mountain stream} image. From the figure, we observe that both SBHex and BBHex outperform EZW (DB2). When compared with JPEG2000, SBHex and BBHex are inferior; the gap is smaller at lower bit rates but widens at higher bit rates with SBHex slightly outperforming BBHex. However, both SBHex and BBHex produce much better quality compared to EZW. 
This result suggests that the incorporation of higher-order wavelets (as is the case in JPEG2000) and appropriate entropy encoding in our schemes can potentially lead to better results compared to JPEG2000. This is a topic for future investigation.

In order to get an overview of the overall performance for all the test images from the datasets under varying bit rates, we plot all the results together in Fig.~\ref{fig:fitting_curve_ezw}. For comparison, the results for JPEG2000 are also shown. We use the SSIM metric for this test so that we can fit a regression curve. The resulting scatter plot has a large spread because it contains SSIM results of a variety of images reconstructed at different bit rates. In order to get a sense of the average performance of the different coding schemes, a Hill function model~\cite{wiki:hillfunction} is fit to the data using non-linear regression. The Hill function is extensively used to model dose-response data in biochemistry and pharmacology~\cite{Goutelle2008TheHE}. 
It  is given by
\begin{equation}
y = f(x; y_{max}, EC_{50}, n) = \frac{y_{max}}{1 + (\frac{EC_{50}}{x})^n}
\label{eq:hilleqn}
\end{equation}
Here, $EC_{50}$ is the input value required to generate the $50\%$ of the maximal response for the dose-response curve. The resulting fits are also shown in 
Fig.~\ref{fig:fitting_curve_ezw}. We observe that on average, both SBHex and BBHex outperform EZW. When compared against each other, SBHex and BBHex produce similar quality results at lower bit rates.

\subsection{Distribution of Symbols}
\begin{figure*}[t!]
  \centering
  \begin{subfigure}[c]{\textwidth}
   
    \raisebox{-.5\height}{\includegraphics[width=\textwidth]
    {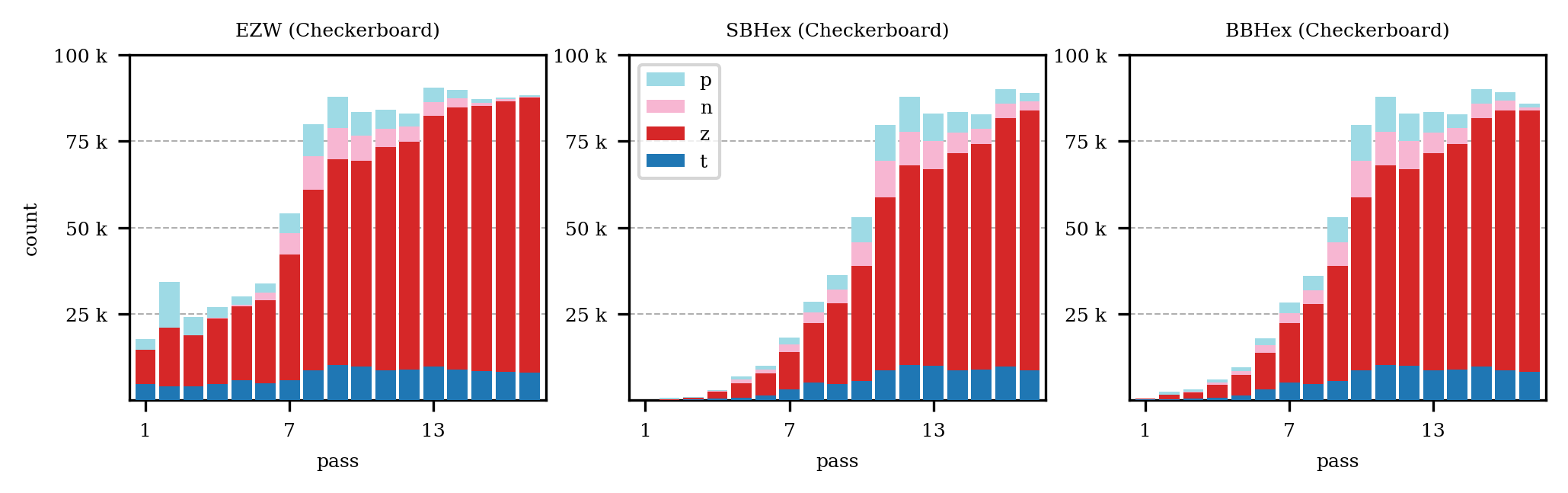}}
\end{subfigure}%
\\
\begin{subfigure}[c]{\textwidth}
    \raisebox{-.5\height}{\includegraphics[width=\textwidth]
    {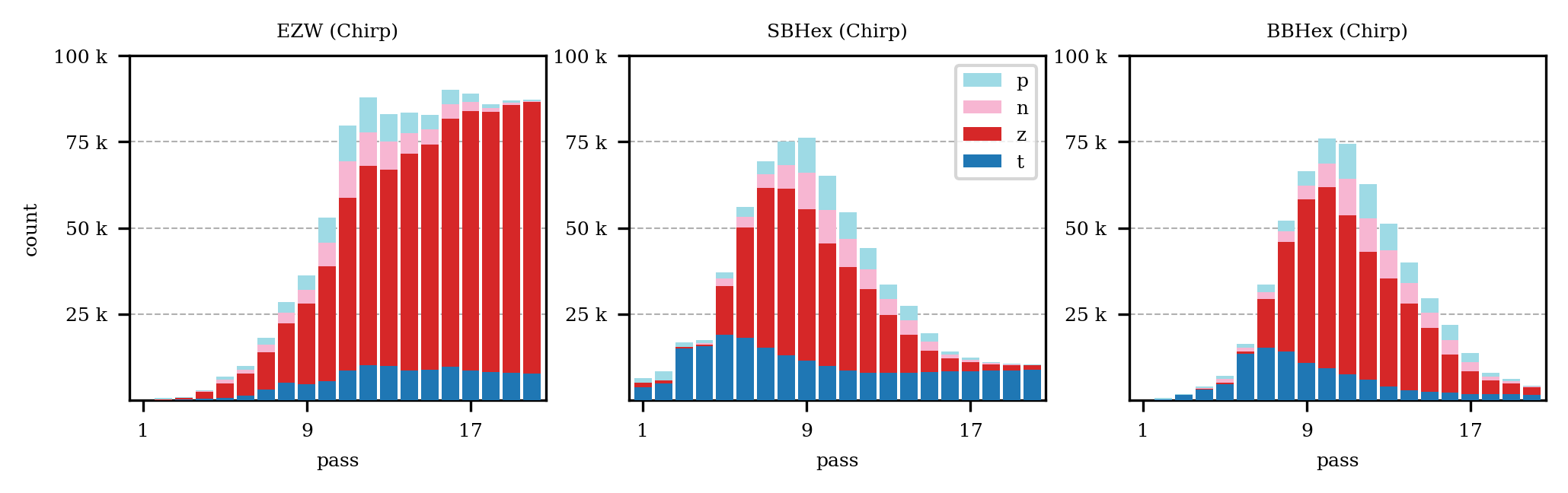}}
\end{subfigure}%
\\
\begin{subfigure}[c]{\textwidth}
    \raisebox{-.5\height}{\includegraphics[width=\textwidth]
    {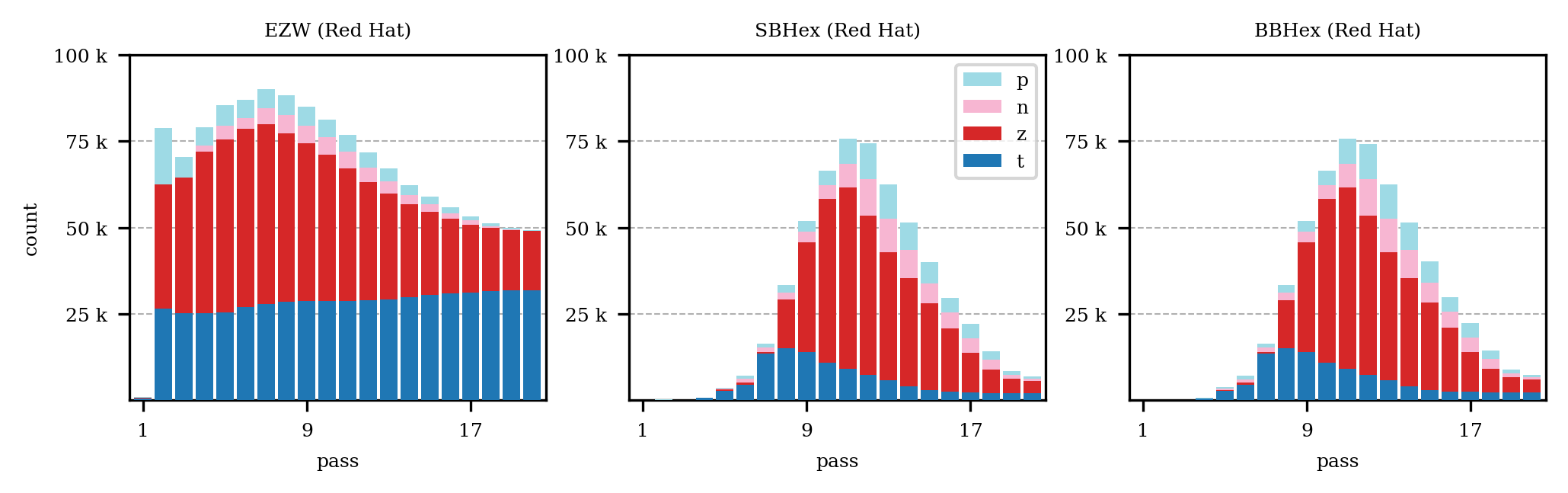}}
\end{subfigure}%

\caption{Distribution of symbols after each pass for EZW (left column), SBHex (middle column) and BBHex (right column), for the checkerboard (top row), chirp (middle row) and Red Hat (bottom row) images.}
\label{fig:significanceMaps}
\end{figure*}
Our proposed coding techniques (SBHex, BBHex) for hexagonal images produce compressed files that are $2-2.5$ times smaller (approximately) than the Cartesian counterparts for all quality levels when BPP $< 4.0$. In order to ascertain why the performance is better at low bit rates compared to EZW, we explored the resulting significance maps after each coding pass and counted the number of 't', 'z', 'n' and 'p' symbols in the code streams. We did this for three different images with different degrees of smoothness. The resulting distributions are charted in Fig.~\ref{fig:significanceMaps}. At first glance, it is apparent that the overall size of the code stream is smaller for SBHex and BBHex compared to EZW. The difference is remarkably pronounced for the chirp and Red Hat images where the SBHex and BBHex distributions exhibit a bell shaped profile. There is significant improvement for the Red Hat image (Fig.~\ref{fig:significanceMaps} - middle row) where SBHex and BBHex capture the information in the first few passes with very few symbols compared to EZW which generates much longer symbol streams. This can be attributed to the isotropy of the hexagonal lattice and wavelet combination which yields sparser wavelet transforms for smooth natural images. The spiral arrangement of sub-bands combined with our hexagonal traversal further enhance the embedding capacity of SBHex and BBHex. When the input image is not so smooth and has discontinuities (as is the case with the synthetic checkerboard image), the difference is not as pronounced. Nevertheless, there is still an advantage compared to EZW (Fig.~\ref{fig:significanceMaps} - top row). 

\begin{figure}[t!]
  \centering
    \includegraphics[scale=0.61]{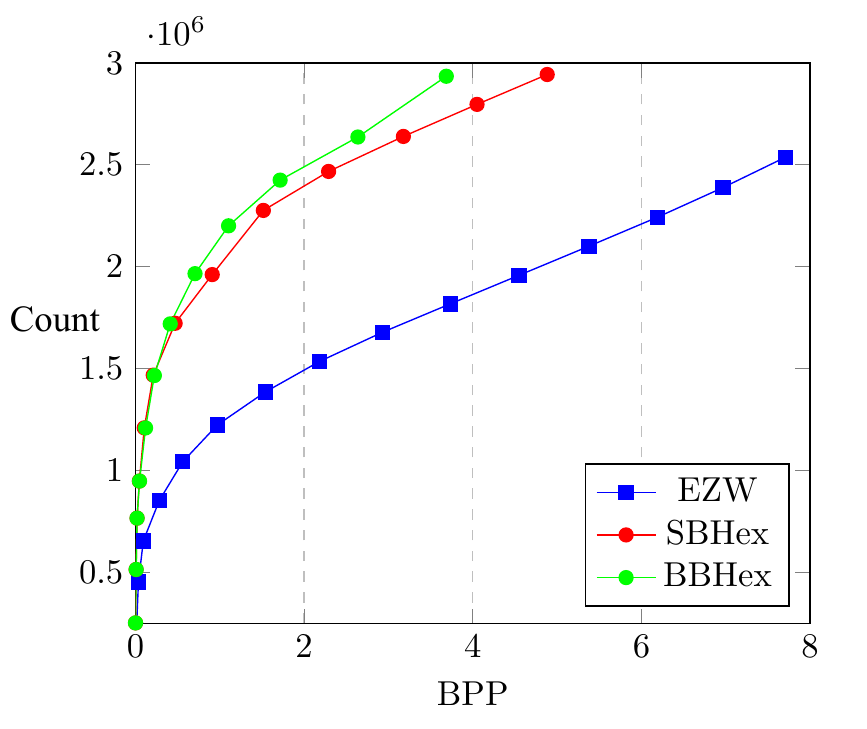}
    \caption{Number of coefficients encoded as zero trees}
    \label{fig:chirp_encoded_zt}
    \caption{Number of coefficients encoded as zero-trees (\emph{ZT}) in different coding schemes for the synthetic ``Chirp" image. Multiresolution analysis is performed to decomposition level 6.}
\end{figure}

In tree-based compression schemes, the higher the number of coefficients encoded as zero-trees, the more the compression. Compression also depends on the size of the tree. Once an entry is found as a zero-tree, then it is not checked further in the magnitude test. From Fig.~\ref{fig:significanceMaps}, a subtle difference between SBHex and BBHex can be seen when it comes to the number of coefficients encoded as zero-trees (`t'). To investigate this further, Fig.~\ref{fig:chirp_encoded_zt} shows that as the bit rate increases, BBHex encodes more coefficients as zero-trees for the chirp image. Both SBHex and BBHex code considerably more coefficients as zero-trees compared to EZW.

In this work, we use straightforward Huffman coding to encode the symbol streams. The results presented in this section suggest that more space saving can be achieved by analyzing the entropy of the code streams. This is a topic for future investigation. 

\subsubsection{Space complexity} Space complexity characterizes the memory requirements. 

\begin{itemize}
    \item There is no memory overhead for the spiral mapping.  
    \item No extra memory cost is associated traversal techniques compared to classical EZW and SPIHT algorithm. The parent-to-child relationship is calculated during run-time. So, no memoization is required. 
    
    \item Memory cost is involved in 2-D index mapping from hexagonal lattice. For $256 \times 256 \times 2$, the size of index map is $511 \times 511$. Considering the extra memory cost of the compression is fair as it assists us to utilize regular convolution routines optimized for the 2-D Cartesian grid.
    
\end{itemize}

\section{Conclusion}
\label{sec:conclusion}

We presented SBHex and BBHex, two tree-based wavelet coding schemes specifically for images sampled on the hexagonal lattice. Both our coding schemes work with images that have been transformed via a second order compactly supported biorthogonal hexagonal wavelet. The coding schemes are designed to exploit the higher compressibility of hexagonally sampled images compared to their Cartesian counterparts. Common to both SBHex and BBHex is a novel hexagonal traversal scheme in the wavelet domain. Our results indicate that SBHex and BBHex outperform EZW across the board. Performance is commensurate with the smoothness of the input images. Smooth natural images show the best results; for low bit rates, our results are almost on par with JPEG2000. 

Hexagonal imaging pipelines are missing a number of key ingredients. This work addresses one of these pieces namely hexagonal image compression. Our results show that if imaging pipelines are modified to acquire images on hexagonal grids, there is a lot to be gained in downstream steps. The images can be readily compressed yielding smaller file sizes which will benefit storage and transmission. 

While we have focused on sub-band coding in this work, our results shed light on prospective research opportunities in the area of hexagonal image compression. 
We are confident that many application areas can be improved by employing these hexagonal coding frameworks. Notwithstanding, we would like to explore more the scalability and extensibility of our coding schemes in future. We conclude by highlighting some of the notable future directions of our work.
\begin{itemize}
    \item Our coding scheme can provide virtually lossless reconstructions, not perfectly lossless. In most applications, this is not a hindrance. One cannot guarantee perfectly lossless image compression via wavelet transform using floating-point filters~\cite{PearlmanWilliamA.WilliamAbraham2013Wic_wavelet_book}. However, perfect reconstruction might be crucial for diagnostic medicine. Perfect reconstruction necessitates integer-to-integer filters for wavelet transformation and a compatible coding scheme. In our coding system, we apply canonical quantization and entropy-coding techniques. Exploring suitable quantization and entropy coding techniques exclusively for hexagonal wavelet filters is a promising direction for future work.
    
    \item Presently, all of our operations are performed with a single CPU thread. However, both SBHex and BBHex can be coded in parallel. GPU implementation of the coding scheme is a point to consider, which will reduce compression time considerably. Moreover, we can split the hexagonal image into tiles. Then each tile of the image can be transformed, encoded, and decoded separately, offering a coding scheme that can be scaled easily to larger image sizes. \\ 
    The extra-padded zeros in our current implementation add a slight overhead. Exploring windowing and memory layout schemes that are better suited to the hexagonal grid is also a subject of future research.
    
    \item In multilevel focus+context visualization~\cite{Hasan16}, a specific region from an image is fetched for decoding and rendering. A future research direction is to decode from a selected area from all sub-bands in the wavelet tree based on a user query.

	\item Higher-order wavelets and wavelets based on the lifting scheme such as loop subdivision wavelets~\cite{bertram2004biorthogonal} need to be investigated as the change of the multiresolution method may alter the performance and quality of compression.

    \item Our processing techniques are entirely image-based and focused on 2-D lattices. However, there is a promising scope of extending the coding scheme to videos and 3D images.
\end{itemize}

\section{Acknowledgements}
We acknowledge the support of the Natural Sciences and Engineering Research Council of Canada (NSERC), [funding reference number RGPIN-2019-05303].

%
%








\bibliographystyle{elsarticle-num}      
\bibliography{citations}   

%
%

\end{document}